%% file: draft_arxiv_v1.tex
\documentclass[prx,aps,twocolumn,groupedaddress,nofootinbib]{revtex4-2} 
\usepackage{graphicx}
\usepackage[nointegrals]{wasysym} %
\usepackage[export]{adjustbox}
\usepackage{amsmath,amsfonts,amssymb,latexsym}
\usepackage{hhline}
\usepackage{bm}
\usepackage{verbatim}
\usepackage{enumitem}
\usepackage{mathrsfs}
\usepackage{slashed}
\usepackage{empheq}
\usepackage{physics}
\input{preamble}

\allowdisplaybreaks

\newcommand{\bs}{\boldsymbol}

\begin{document}


\title{Metallogenic quantum criticality: \\ Fermi surface nucleation at transitions between gapped phases}

\author{Zhengyan Darius Shi}%
\email{zhengyanshi@stanford.edu}
\affiliation{
Leinweber Institute for Theoretical Physics, Stanford University, Stanford, California 94305, USA}

\date{\today}

\begin{abstract}
    Continuous quantum phase transitions in interacting many-body systems exhibit universal phenomena that are largely independent of microscopic details. An organizing principle distilled from canonical examples such as the magnetic transition in Ising models is that the low-energy physics of transitions between gapped phases is governed by scale-invariant quantum field theories with gapless degrees of freedom at isolated points in momentum space. In this paper, we show that this principle can fail in a radical way: an entire gapless Fermi surface can emerge at a quantum critical point separating two fully gapped insulators. These exotic phase transitions are dubbed \textit{metallogenic quantum critical points} (MGQCP). We construct a controlled family of effective field theories for MGQCPs between distinct fractional Chern insulators and propose a periodically modulated quantum Hall bilayer as a minimal microscopic setting. The critical theory exhibits an unusual mixture of physical signatures: while transport and compressibility follow scaling laws compatible with a conformal field theory, the electron spectral function develops a power-law non-analyticity at the Fermi momentum, characteristic of non-Fermi-liquid metals. We discuss how the concept of MGQCP may shed new light on critical phenomena in moire materials with tunable Chern bands. More generally, our results point to a broader class of quantum critical phenomena in which extended manifolds of gapless excitations emerge in momentum space despite being absent in both neighboring phases. 
\end{abstract}

\maketitle

\section{Introduction}

Continuous quantum phase transitions occupy a privileged place in the study of quantum matter~\cite{Sondhi1997_QCPRMP,Sachdev1999_QCParticle}. The divergence of correlation lengths at such phase transitions erases the system's memory of microscopic details and reveals universal scaling laws that constrain observables throughout the quantum-critical regime. This emergence of universal structure makes quantum critical points natural anchors for complex phase diagrams.

An important class of quantum critical points arises at transitions between gapped phases. Over the past few decades, tremendous progress has been achieved in classifying and characterizing gapped phases, vastly generalizing the conventional Landau classification based on broken symmetries~\cite{Witten1989_CSTQFT,Wen1990,Wen1992,Levin2004,Kitaev2005,Schnyder2008,Kitaev2009,Chen2010a,Chen2011,
Lan2016,Freed2016}. In parallel, a rich variety of non-Landau critical phenomena between gapped phases has also been discovered. Prominent examples include Chern-Simons-matter critical points between topologically ordered phases~\cite{Lee1991_QHinsulator_QCP,Kivelson1992_globalFQHQCP,Wen1993_topoQCP,Chen1993_anyonMotttransition,Wen1999,Barkeshli2010_topoQCP,Barkeshli2012_FCISF_QCP,Grover2012_BIQHtransition,Lee2018_QED3QCP}, deconfined critical points between phases with incompatible broken symmetries~\cite{Senthil2003_DQCPshort,Senthil2003_DQCPlong,Vishwanath2003_VBSVBS,Vishwanath2012_surfaceDQCP,Wang2017_DQCPduality,Senthil2023_DQCPreview}, and distinct universality classes separating identical neighboring phases~\cite{Bi2018_adventure,Bi2019_LBL,Prakash2023_1Dmultiversal_unnecessaryQCP}.
These discoveries have dramatically broadened the landscape of quantum criticality, yet they respect a common structural principle: every transition between gapped phases tuned by a single parameter is described by a scale-invariant quantum field theory, with gapless degrees of freedom concentrated near a finite set of points in momentum space. Such theories arise naturally as fixed points of renormalization-group (RG) flows and provide the basis for our most powerful analytical and numerical approaches to critical phenomena~\cite{Wilson1971_RGI,Wilson_1971_RGII,Wilson_1972_eps,Poland2018_bootstrap_review,He2026_fuzzyreview}.

In this work, we show that this basic principle can fail in a radical way. Specifically, we construct a family of \textit{metallogenic quantum critical points} (MGQCPs) where an entire critical Fermi surface with a codimension-1 (in momentum-space) manifold of gapless excitations is born at the transition between two gapped insulators. The family is labeled by an integer $p$ and is analytically controlled in the large $p$ limit. The existence of a Fermi surface supplies a momentum scale $k_F$ which violates conventional scale invariance at criticality (see Refs.~\cite{Polchinski1992_FSRG,Shankar1993_FSRG,Ma2023_FLfRG,Kukreja2024_FSprojectiveRG,Kukreja2026_FSprojectiveRG} for attempts to formulate a generalized notion of RG and scale-invariance for systems with Fermi surfaces). We emphasize that MGQCPs are conceptually different from existing ``metallic'' QCPs, where a Fermi surface is already present in at least one neighboring phase~\cite{Hertz1976_metalQCP,Millis1993_metalQCP,Senthil2004_HFQCP,Senthil2008_continuousMott,Lee2017_NFLreview,Zou2020_DMITU2,Zou2020_DMITbilayer}. 

\begin{figure}
    \centering
    \includegraphics[width=\linewidth]{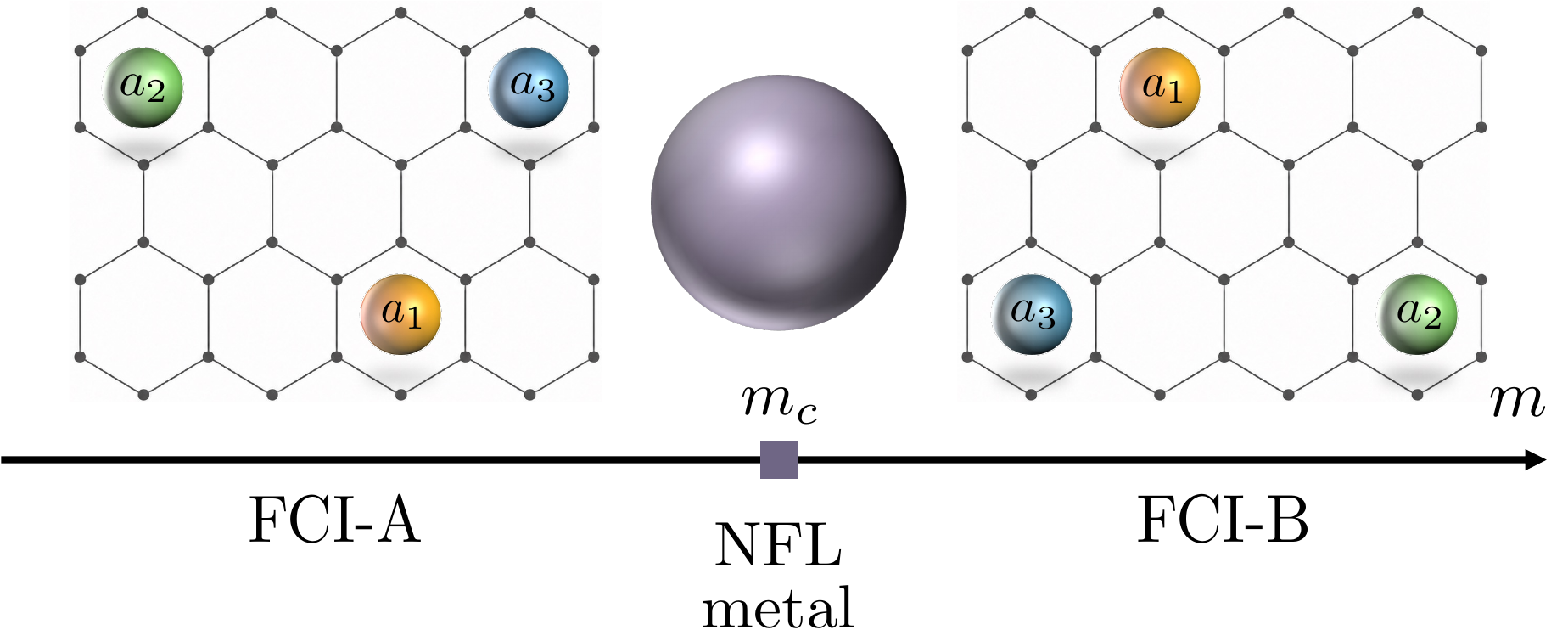}
    \caption{Schematic phase diagram in which a critical non-Fermi liquid (NFL) metal with a gapless Fermi surface emerges at the phase transition between two fully gapped fractional Chern insulators (FCIs) with Abelian anyons $a_i$.}
    \label{fig:phase_diagram}
\end{figure}

The emergence of a critical Fermi surface at the MGQCP leads to striking physical predictions. In the single-electron spectral function $A_c(\bs{k}, \omega)$, we find a non-analyticity across the Fermi surface, which is the defining feature of electronic non-Fermi liquids. In thermodynamics, the system is incompressible despite having a non-Fermi liquid specific heat $C(T) \sim T \log (\Lambda/T)$. In contrast, the frequency-dependent resistivity $\rho(\omega, T)$ is largely insensitive to the Fermi surface and takes the scaling form in an ordinary conformal field theory. 

\begin{figure}
    \centering
    \includegraphics[width=0.8\linewidth]{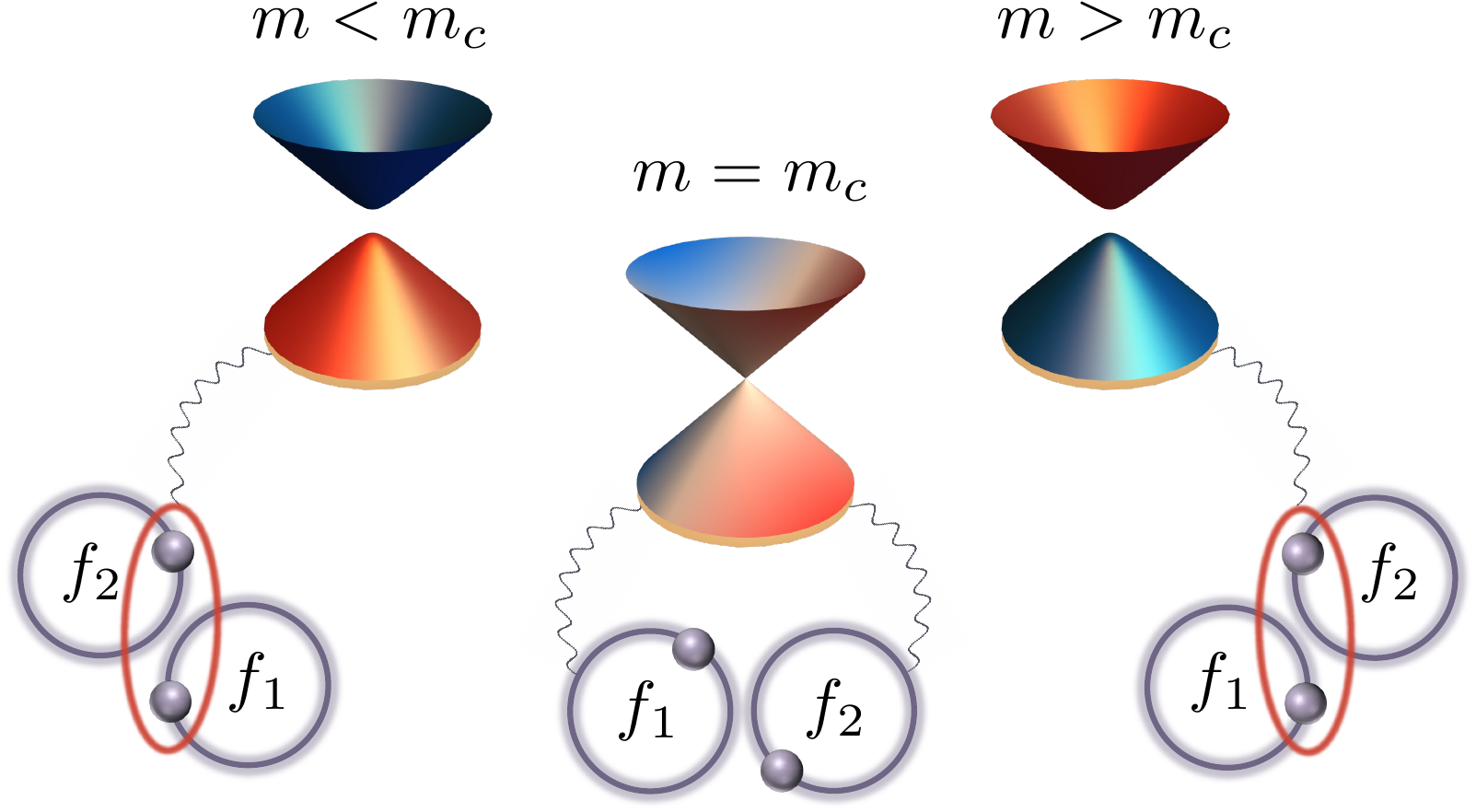}
    \caption{Physical mechanism for a \textit{metallogenic quantum critical point}. In the vicinity of the critical point, the system decomposes into a QED$_3$ (2+1D quantum electrodynamics) sector and two Fermi surfaces $f_1, f_2$ coupled to the same set of emergent $U(1)$ gauge fields. For either sign of the Dirac mass in QED$_3$, gauge fluctuations induce a pairing instability between $f_1$ and $f_2$, leading to fully gapped phases. Precisely at criticality, gapless Dirac fermions modify the gauge field dynamics and stabilize the Fermi surfaces. }
    \label{fig:intuition}
\end{figure}

The mechanism underlying MGQCPs can be understood intuitively by approaching them from the finite-temperature regime. The minimal MGQCP we construct involves two-component electrons in a partially filled Chern band. At intermediate temperatures, the system realizes a transition between two types of generalized composite Fermi liquids (gCFLs) with two composite fermion (CF) Fermi surfaces coupled to emergent $U(1)$ gauge fields~\cite{Jain1989_CFframework,Lopez1991_CSGL,Kalmeyer1992_CFL,Halperin1993_HLR}. Away from the critical point, gauge fluctuations induce a strong pairing instability that ultimately gaps out the Fermi surfaces at low temperature, much like in Landau-level bilayer CFLs~\cite{Milovanovic2007_CFpairing,Moller2008_CFpairing,Moller2009_CFpairing,Milovanovic2015_CFpairing,Sodemann2017_CFLbilayer,Isobe2016_CFLbilayer}. The resulting electronic ground states are two distinct gapped fractional Chern insulators (FCIs)~\cite{Sun2011_FCI,Sheng2011_FCI,Regnault2011_FCI,Tang2011_FQAH,Wang2011_FQAH,Neupert2010_FQAH,Bergholtz2013_FCIreview,Parameswaran2013_FCIreview}.\footnote{This mechanism for FCIs from pairing of generalized CFs should be reminiscent of the celebrated Moore–Read quantum Hall state, which is a $p+ip$ paired state of conventional CFs.} The remarkable phenomenon, as illustrated in Fig.~\ref{fig:intuition}, is that quantum criticality suppresses the gauge-mediated pairing tendency, thereby stabilizing the CF Fermi surfaces precisely at the transition point.

Going beyond transitions between gapped phases, our construction suggests the general principle that gapless structures emerging at a quantum critical point need not resemble the elementary excitations of either neighboring phase and, in particular, may possess a momentum-space dimensionality absent throughout the surrounding phase diagram. These criticality-stabilized gapless structures can produce physical signatures sharply distinct from scale-invariant critical points, as we elaborate upon in the rest of this paper.

\section{Model: quantum Hall bilayer with tunable periodic potential}\label{sec:model}

The minimal candidate microscopic realization of MGQCPs is a lattice quantum Hall bilayer system, where the fermion $c_i$ in each layer sees a large magnetic field $B_i = \nabla \times \bs{A}_i$ and a lattice periodic potential $W_i(\bs{r})$. In the simplest setup, we choose $B_1 = B_2 = B$, $W_1 = W_2 = W$, and arrange the lattice so that each unit cell encloses exactly $4\pi$ magnetic flux. The non-interacting Lagrangian thus enjoys an internal symmetry $G_{\rm int} = \left[U(1)_1 \times U(1)_2\right] \rtimes \mathbb{Z}_2$, where $U(1)_i$ corresponds to charge conservation in layer $i$, and $\mathbb{Z}_2$ exchanges the two layers. The dominant interaction is a $G_{\rm int}$-symmetric density-density repulsion
\begin{equation*}
    H_{\rm int} = \frac{1}{2} \int_{\bs{q}} \begin{pmatrix}
        \rho_1(\bs{q}) & \rho_2(\bs{q})
    \end{pmatrix} \begin{pmatrix}
        V(\bs{q}) & U(\bs{q}) \\ U(\bs{q}) & V(\bs{q}) 
    \end{pmatrix} \begin{pmatrix}
        \rho_1(-\bs{q}) \\ \rho_2(-\bs{q})
    \end{pmatrix} \,. 
\end{equation*}
Here, $\rho_i = c^{\dagger}_i c_i$ is the density operator in layer $i$ and $V(\bs{r}) = V_0/|\bs{r}|^{1+\epsilon}, U(\bs{r}) = V_0/(|\bs{r}|^2+d^2)^{(1+\epsilon)/2}$ is a one-parameter family of intralayer and interlayer interaction potentials, with $d$ denoting the layer separation. The constant $\epsilon \geq 0$ characterizes the strength of screening, and vanishes for unscreened Coulomb interactions. Consistent with $G_{\rm int}$, we choose $\rho_i$ such that both layers have the same Landau level filling $\nu$. We work with $\nu = 1/3$ for the simplest constructions and comment on generalizations to other Jain fractions $\nu = p/(2p+1)$ in Sec.~\ref{sec:discussion}. 

In what follows, we construct a low-energy effective theory for a continuous transition between two gapped FCIs, A and B, compatible with the microscopic symmetries and anomalies associated with the above setup. We will show that the quantum critical point has a single relevant deformation and therefore describes a codimension-1 region in the global phase diagram. This argument establishes the existence of MGQCPs in principle without committing to a specific path of Hamiltonians.

Complementary to the effective theory, we also propose a possible microscopic realization of the MGQCP in a two-parameter family of models with periodic potentials of the form
\begin{equation}\label{eq:periodic_potential}
    \begin{aligned}
    W(\bs{r}) = &\,\,W \cos (2\pi x) \cos (2\pi y) \\
    &+ \alpha \, W \left[\cos (2\pi x) + \cos (2\pi y)\right] \,,
    \end{aligned}
\end{equation}
where distance units are chosen such that the lattice constant $a = 1$. We give analytic arguments in support of this proposal and defer a more thorough numerical study to future work. 

\section{Generalized composite fermions}

A powerful theoretical framework for describing fractional quantum Hall phases is the parton construction, in which the microscopic electron is splintered into several constituent fermionic/bosonic partons which are glued together by emergent dynamical gauge fields~\cite{Lee2004_dopeMott}. Through this construction, complex fractionalized phases of microscopic electrons reduce to simple mean-field states of partons with static gauge fluxes. 

In the quantum Hall bilayer, we employ a specific parton construction (recently utilized in Ref.~\cite{Han2025_anyonexciton}) which is a direct generalization of the celebrated composite fermion (CF) construction in single layer quantum Hall systems~\cite{Jain1989_CFframework,Lopez1991_CSGL,Kalmeyer1992_CFL,Halperin1993_HLR}. Specifically, we write $c_i = f_i \Phi_i$, where $f_i$ is a fermion and $\Phi_i$ is a hard-core boson. For each $i$, this decomposition introduces a $U(1)$ gauge redundancy under which $f_i(\bs{r}) \rightarrow f_i(\bs{r}) e^{-i \theta_i(\bs{r})}, \Phi_i(\bs{r}) \rightarrow \Phi_i(\bs{r}) e^{i \theta_i(\bs{r})}$. Therefore, the low energy effective Lagrangian involves $f_i, \Phi_i$ and a pair of dynamical $U(1)$ gauge fields $\{\alpha_i\}$
\begin{equation}
    L_{\rm eff} = \sum_{i=1}^2 \left( L[f_i, A_i - \alpha_i] + L[\Phi_i, \alpha_i] \right) \,, 
\end{equation}
where $A_i$ is the background gauge field associated with the layer $U(1)_i$ symmetry and the $\mathbb{Z}_2$ symmetry acts by exchanging $f_1 \leftrightarrow f_2, \Phi_1 \leftrightarrow \Phi_2$. 

In the spirit of the CF construction, we choose a mean-field in which $\nabla \times \bs{\alpha_i} = \nabla \times \bs{A_i} = B$ so that each $f_i$ sees zero magnetic flux on average and can form a Fermi surface. On the other hand, the hard-core bosons $\{\Phi_i\}$ see the total magnetic flux and can form a gapped bosonic bilayer quantum Hall state. Without accounting for gauge fluctuations, every choice of bosonic FCI for $\{\Phi_i\}$ leads to a bilayer generalized composite Fermi liquid (gCFL) at arbitrary layer filling $\nu_1 = \nu_2 = \nu$.\footnote{Similar gCFL phases have been considered in Refs.~\cite{Zou2020_DMITU2,Zou2020_DMITbilayer}.} In the rest of this section, we will proceed in two steps. First, we construct two specific bosonic FCIs for $\{\Phi_i\}$ in Sec.~\ref{subsec:boson_bilayer} and argue that the tunable periodic potential in \eqref{eq:periodic_potential} can drive a continuous transition between them. Then, we couple each bosonic FCI to the Fermi surfaces of $\{f_i\}$ in Sec.~\ref{subsec:boson_fermion_coupling} and show that gauge fluctuations drive a pairing instability. These results leave open the fate of the Fermi surfaces precisely at the critical point, which we address in Sec.~\ref{sec:crit_stabilized_FS}.

\subsection{Bosonic transition underlying the MGQCP}\label{subsec:boson_bilayer}

We begin by analyzing the bosonic bilayer $\{\Phi_i\}$ at Landau level filling $\nu_1 = \nu_2 = 1/3$, subject to a periodic potential of the form \eqref{eq:periodic_potential}. Throughout the analysis, we assume that both $W$ and $V$ (the Coulomb interaction scale) are much smaller than the cyclotron gap $\omega_c$ of $\Phi_i$. 

When $W = 0$, our system reduces to the standard bosonic quantum Hall bilayer at $\nu_i = 1/3$. At sufficiently small interlayer distance $d$, large-scale DMRG studies provide strong evidence that the ground state is a symmetry-preserving bosonic Jain state with three anyons~\cite{Geraedts2017_onethird_Bbilayer}, which we will refer to as bFCI-A. The corresponding topological quantum field theory (TQFT) is 
\begin{equation}\label{eq:L_bFCIA}
    L^{(A)}_{\rm bFCI} = 3 \mathrm{CS}[a,g] - \frac{1}{2\pi} a d (\alpha_1 + \alpha_2) + \sum_{i=1}^2 \frac{1}{4\pi} \alpha_i d \alpha_i \,, 
\end{equation}
where $\mathrm{CS}[a,g] \equiv \frac{1}{4\pi} a da + 2 \mathrm{CS}_g$ is the standard response theory of a $C = 1$ Chern insulator to a $U(1)$ gauge field $a$ (technically a $\mathrm{spin}_{\mathbb{C}}$ connection) and a background metric $g$. In Appendix~\ref{app:microscopic}, we show that this TQFT is equivalent to the familiar Halperin (221) state, which realizes the \textit{unique minimal} $G_{\rm int}$-symmetric topological order with $c_- = 2$, Hall conductance $\sigma_{xy} = 2/3$, and counterflow Hall conductance $\sigma^{\rm cf}_{xy} = 2$~\cite{Cheng2025_orderingFQH}. 

Next, we increase the periodic potential strength $W$ while holding the dimensionless constant $\alpha$ fixed. For $\alpha > 0$, the potential splits the lowest LL into two subbands, the lower of which has Chern number $C = 1$~\cite{Lee2018_QED3QCP}. At sufficiently large $W$ (still much smaller than the cyclotron gap), the system is thus well-described by a pair of bosons $\Phi_i$, each at $2/3$ filling of a $C = 1$ band. By tuning the value of $\alpha$, we can further flatten the $C = 1$ band to favor the ``bosonic particle-hole conjugate" of bFCI-A, which can be regarded as a boson integer quantum Hall state stacked with the time-reversal of bFCI-A. This new state, which we refer to as bFCI-B, has the smallest anyon count consistent with its Hall conductance $\sigma_{xy} = 4/3$, and is indeed realized by two-component bosons at total Landau level filling $4/3$~\cite{Geraedts2017_onethird_Bbilayer}. Guided by the principle of minimality, we postulate that increasing $W$ drives a direct transition from bFCI-A to bFCI-B. In Appendix~\ref{app:microscopic}, we show that the simplest Lagrangian of bFCI-B takes the form
\begin{equation}\label{eq:L_bFCIB}
    \begin{aligned}
        L^{(B)}_{\rm bFCI} &= - 3 \mathrm{CS}[a,g] + \frac{2}{2\pi} a d (\alpha_1 + \alpha_2) - \frac{2}{4\pi} \sum_{i=1}^2 \alpha_i d \alpha_i \,.
    \end{aligned}
\end{equation}
Quantized invariants of both phases are shown in Tab.~\ref{tab:topological_bFCIs}. 
\begin{table}
    \centering
    \begin{ruledtabular}
    \begin{tabular}{ccccc}
        Phase & Topological order & $\sigma_{xy}$ & $\sigma^{\rm cf}_{xy}$ & $c_-$ \\
        bFCI-A & (221) & $2/3$ & $2$ & $2$ \\ 
        bFCI-B & $\overline{(221)} \times \mathrm{bIQH}$ & $4/3$ & $-4$ & $-2$ 
    \end{tabular}
    \end{ruledtabular}
    \caption{Summary of topological properties on two sides of the bosonic transition. (221) is the Halperin state with K-matrix $K_A = 2 I_{2 \times 2} + \sigma_x$ and (221) is its time-reversal conjugate. bIQH is the boson integer quantum Hall state. $\sigma_{xy}, \sigma^{\rm cf}_{xy}$ are the Hall conductivities in the charge and counterflow channels and $c_-$ is the chiral central charge.}
    \label{tab:topological_bFCIs}
\end{table}

To describe the phase transition from bFCI-A to bFCI-B, it is useful to obtain both phases through a parton construction $\Phi_i = \chi_i \psi$ where $\chi_i, \psi$ are fermionic partons glued together by a $U(1)$ gauge field $a$\footnote{Technically, $a$ is a $\mathrm{spin}_{\mathbb{C}}$ connection. This subtlety is not essential for this paper.}
\begin{equation}
    \sum_{i=1}^2 L[\Phi_i, \alpha_i] = \sum_{i=1}^2 L[\chi_i, \alpha_i - a] + L[\psi, a] \,. 
\end{equation}
The equations of motion for $a$ impose the density constraint $\rho_{\chi_1} = \rho_{\chi_2} = \rho_{\psi}/2$ so that $\chi_i$ is at lattice filling $2/3$ and $\psi$ is at lattice filling $4/3$. To obtain gapped bosonic FCIs, we choose a mean-field $\nabla \times \bs{a} = 2 \nabla \times \bs{\alpha}_i/3$ such that $\chi_1, \chi_2$ each sees $4\pi/3$ flux and $\psi$ sees $8\pi/3$ flux per unit cell. Solving the corresponding Hofstadter problem, we see that $\chi_i, \psi$ can form Chern insulators with Chern numbers $C_{\chi_i}, C_\psi = 1$ mod 3. 

If we keep $C_{\psi} = 1$ fixed and choose $C_{\chi_i} = 1$ or $C_{\chi_i} = -2$, integrating out all gapped partons gives the TQFT in \eqref{eq:L_bFCIA} or \eqref{eq:L_bFCIB} respectively. Thus, the transition between bFCI-A and bFCI-B maps to a band-inversion transition for each $\chi_i$, with three degenerate gapless Dirac cones in momentum space related by the action of lattice translations~\cite{Lee2018_QED3QCP}.

Using a regularization in which $L[\Psi, A]$ describes a gapless Dirac fermion with $\sigma_{xy} = 0$, we arrive at the following critical theory between bFCI-A and bFCI-B
\begin{equation}\label{eq:bFCI_CFT}
    \begin{aligned}
    &L_{\rm crit, \Phi} = \sum_{i=1}^2 \sum_{n = 1}^3 L[\Psi_{i, n}, \alpha_i - a] + \frac{1}{4\pi} a da \\
    &- \frac{1}{8\pi} (\alpha_1 - a) d (\alpha_1 - a) - \frac{1}{8\pi} (\alpha_2 - a) d (\alpha_2 - a)  \,. 
    \end{aligned}
\end{equation}
Since $\alpha_i$ are background fields for the $\{\Phi_i\}$ sector, the only dynamical field is $a$, which has no Chern-Simons level at the critical point. The transition between bFCI-A and bFCI-B is thus $N_f = 6$ QED$_3$, which is likely conformal according to bootstrap and large-$N_f$ calculations~\cite{Chester2016_bootstrapQED,Li2018_bootstrap}. We conclude that the bosonic sector of our theory is an ordinary topological transition between two bFCIs described by a conformal field theory (CFT). 

\subsection{Gapped fermionic FCIs from pairing of generalized composite fermions}\label{subsec:boson_fermion_coupling}

We now couple the bosonic sector back to the Fermi surfaces of $\{f_i\}$. On both sides of the bosonic transition, the bosonic FCIs implement a flux attachment constraint that relates the CF density $\rho_{f_i}$ to linear combinations of $\nabla \times \bs{\alpha}_i$. Through flux attachment, density-density interactions of CFs modify the gauge field dynamics, which in turn backreact on the CF Fermi surfaces. 

In a single-layer CFL, gauge fluctuations tend to suppress pairing and the mean-field CF Fermi surface remains stable down to zero temperature~\cite{Bonesteel1998_CFLpairing,Metlitski2014_NFLpairing}. In contrast, analytical and numerical studies show that gauge fluctuations in bilayer CFLs generically induce a strong interlayer pairing instability~\cite{Milovanovic2007_CFpairing,Moller2008_CFpairing,Moller2009_CFpairing,Milovanovic2015_CFpairing,Sodemann2017_CFLbilayer,Isobe2016_CFLbilayer}. For inversion-symmetric Fermi surfaces, such an instability gaps out the entire Fermi surface of composite fermions and leads to a fully gapped fractional quantum Hall phase.\footnote{The inversion-asymmetric case leads to more exotic gapless \textit{composite Bogoliubov Fermi liquids} as recently studied in Ref.~\cite{Shi2026_CBFL}.}

An analogous pairing instability also arises in the specific gCFLs that we have constructed
\begin{equation}
    L^{(A/B)}_{\rm gCFL} = \sum_{i=1}^2 L[f_i, A_i - \alpha_i] + L_{\rm bFCI}^{(A/B)} \,. 
\end{equation}
Relegating details to Appendix~\ref{app:CFL_pairing}, we find that flux attachment transforms the Coulomb interaction into
\begin{equation*}
    \begin{aligned}
    L^{(A)}_{\rm int} &= \frac{V+U}{72\pi^2} (\nabla \times \bs{\alpha}_+)^2 + \frac{V-U}{8\pi^2} (\nabla \times \bs{\alpha}_-)^2 \,, \\
    L^{(B)}_{\rm int} &= \frac{V+U}{18 \pi^2} (\nabla \times \bs{\alpha}_+)^2 + \frac{V-U}{2\pi^2} (\nabla \times \bs{\alpha}_-)^2 \,,
    \end{aligned}
\end{equation*}
where all momentum arguments are implicit and $\alpha_{\pm} = (\alpha_1 \pm \alpha_2)/\sqrt{2}$. Recalling the explicit form of $V(\bs{q})$ and $U(\bs{q})$, we immediately see that the antisymmetric gauge field $\alpha_-$ has a softened $|\bs{q}|^2$ kinetic term, while the symmetric gauge field $\alpha_+$ retains a $|\bs{q}|^{1+\epsilon}$ kinetic term. 

Following Ref.~\cite{Sodemann2017_CFLbilayer,Zou2020_DMITbilayer,Zou2020_DMITU2}, we obtain the one-loop RG flow of dimensionless gauge couplings $g_{\pm}$ and BCS coupling $V_{\rm BCS, m}$ in angular momentum channel $m$
\begin{equation}
    \begin{aligned}
        \frac{d g_+}{dl} &= (\epsilon/2 - g_+ - g_-) g_+ \,, \\
        \frac{d g_-}{dl} &= (1/2 - g_+ - g_-) g_- \,, \\
        \frac{d V_{{\rm BCS}, m}}{dl} &= g_+ - g_- - V_{{\rm BCS}, m}^2 \,. 
    \end{aligned}
\end{equation}
In the physical range $0 \leq \epsilon < 1$, the only stable fixed point for $g_{\pm}$ is $(g_+, g_-) = (0, 1/2)$. Plugging these values into the BCS flow, we find that $V_{\rm BCS}$ runs towards $-\infty$, inducing a robust pairing instability on both sides of the phase transition. Note that the pairing susceptibility is independent of $m$ and the specific pairing channel realized in any microscopic model is fixed by non-universal details~\cite{Metlitski2014_NFLpairing}. We therefore proceed with a general angular momentum $l_{A/B}$, so that the topological quantum field theories (TQFTs) for phase A and B reduce to
\begin{equation}\label{eq:fermionic_FCI_TQFT}
    \begin{aligned}
    &L^{(A)}_{l_A} = \frac{1}{4\pi} \alpha_1 d \alpha_1 + \frac{1}{4\pi} (\alpha_1 - A_1 - A_2) d (\alpha_1 - A_1 - A_2) \\
    &+ l_A \mathrm{CS}[A_1 - \alpha_1, g] + 3 \mathrm{CS}[a,g] - \frac{1}{2\pi} a d (A_1 + A_2) \,, \\
    &L^{(B)}_{l_B} = - \frac{2}{4\pi} \alpha_1 d \alpha_1 - \frac{2}{4\pi} (\alpha_1 - A_1 - A_2) d (\alpha_1 - A_1 - A_2) \\
    &+ l_B \mathrm{CS}[A_1 - \alpha_1, g] - 3\mathrm{CS}[a,g] + \frac{2}{2\pi} a d (A_1 + A_2) \,.
    \end{aligned}
\end{equation}
The possible gapped FCIs and their basic topological properties are summarized in Table.~\ref{tab:gapped_FCIs}. Note that $l_A = -1$ maps to the Halperin (112) state, which is indeed the DMRG ground state of a fermionic bilayer quantum Hall system at small interlayer distance $d \ll l_{\rm mag}$ and vanishing periodic potential $W = 0$~\cite{Geraedts2015_onethird_Fbilayer}. This matching provides support for the relevance of our parton construction to the microscopic model, at least in the small-$W$ regime. 
\begin{table}
    \centering
    \begin{ruledtabular}
    \begin{tabular}{ccccc}
        $l$ & $K$-matrix & $\sigma_{xy}$ & $\sigma^{\rm cf}_{xy}$ & $c_-$ \\
        $l_A$ & $\mathrm{diag}(l_A+2,3)$ & 2/3 & $2l_A/(l_A+2)$ & $l_A + 2 - \sgn(l_A + 2)$ \\
        $l_B$ & $\mathrm{diag}(l_B-4, -3)$ & 4/3 & $4l_B/(4-l_B)$ & $l_B - 2 - \sgn(l_B - 4)$
    \end{tabular}
    \end{ruledtabular}
    \caption{Summary of topological properties on two sides of the MGQCP derived in Appendix~\ref{app:microscopic}. In the $2 \times 2$ diagonal $K$-matrix, the first/second component is the level of a $U(1)$ gauge field/$\mathrm{spin}_{\mathbb{C}}$ connection.}
    \label{tab:gapped_FCIs}
\end{table}

\section{Criticality-stabilized Fermi surface}\label{sec:crit_stabilized_FS}

The preceding analysis establishes that within the parton construction $c_i = f_i \Phi_i$, the formation of bFCI-A or bFCI-B in the $\{\Phi_i\}$ sector induces a strong pairing instability between the generalized CFs $f_i$. These paired states of generalized CFs correspond to gapped electronic FCIs, whose precise anyon content depends on the choice of pairing channel, as illustrated in Tab.~\ref{tab:gapped_FCIs}. 

Given that $f_1$ and $f_2$ suffer from a pairing instability on both sides of the bosonic transition, one may conjecture that interlayer pairing persists even at the critical point, where the $\{\Phi_i\}$ sector is described by the CFT in \eqref{eq:bFCI_CFT}. If this is the case, the $\{f_i\}$ Fermi surfaces are gapped throughout the phase diagram and do not affect low energy physics. Surprisingly, we find that the opposite is true: gapless $\{\Phi_i\}$ fluctuations of the CFT in fact destroy the pairing instability and stabilize the Fermi surfaces of $\{f_i\}$, leading to a non-Fermi liquid critical point.

To analyze this stabilization, we combine \eqref{eq:bFCI_CFT} and the $\{f_i\}$ Fermi surfaces to obtain the complete critical Lagrangian in Euclidean signature
\begin{equation}\label{eq:full_crit_Lagrangian}
    L^E_{\rm crit} = \sum_{i=1}^2 L^E[f_i, A_i - \alpha_i] + L^E_{\rm crit, \Phi}[\alpha_i] \,. 
\end{equation}
The stability of $f_i$ depends crucially on the renormalized propagator of the gauge field $\alpha_i$. Integrating out the gapless Dirac fermions in \eqref{eq:bFCI_CFT}, we see that 
\begin{equation}\label{eq:Dirac_contribution_alpha}
    \begin{aligned}
    L^E_{\rm crit, \Phi}[\alpha_i] &= \frac{1}{2} \alpha_+^{\mu} \left(\frac{q}{g_+} P_{\mu\nu}\right) \alpha_+^{\nu} + \frac{i}{8\pi} \alpha_+ d \alpha_+ \\
    &+ \frac{1}{2} \alpha_-^{\mu} \left(\frac{q}{g_-} P_{\mu\nu}\right) \alpha_-^{\nu} - \frac{i}{8\pi} \alpha_- d \alpha_- \,,
    \end{aligned}
\end{equation}
where $g_+^{-1}, g_-^{-1}$ are conductivities of the $N_f = 6$ QED$_3$ theory with respect to the gauge fields $\alpha_+$ and $\alpha_-$. In terms of the Dirac fermion fields, $g_+^{-1}, g_-^{-1}$ are related to correlation functions of the topological current $J_+^{\mu}$ and the flavor current $J_-^{\mu}$ respectively
\begin{equation}
    \begin{aligned}
    J_+^{\mu} &\equiv \frac{1}{2\sqrt{2}\pi} \epsilon^{\mu\nu\lambda}\partial_{\nu} a_{\lambda} \,, \\
    J_-^{\mu} &\equiv \sum_{s = 1}^2 \sum_{n=1}^3 \frac{(-1)^{s-1}}{\sqrt{2}} \bar \Psi_{s,n} \gamma^{\mu} \Psi_{s,n} \,. 
    \end{aligned}
\end{equation}
In addition to the CFT sector, the $\alpha_i$ kinetic term also gets renormalized by Fermi surface fluctuations of $\{f_i\}$. Treating Fermi surface fluctuations at the level of random phase approximation, we find the Euclidean Lagrangian
\begin{equation*}
    \begin{aligned}
    L^E_{\rm FS}[\alpha_i] &\approx \sum_{s = \pm} \frac{1}{2} K_{s, 00} |\alpha_s^0|^2 + \frac{1}{2} \left[R \, |\bs{q}|^2 + \gamma |\omega|/|\bs{q}|\right] |\alpha_s^T|^2 \,,
    \end{aligned}
\end{equation*}
where $K_{\pm, 00}$ are interaction-renormalized correlation functions for $\rho_{f_1} \pm \rho_{f_2}$ and $R, \gamma$ are positive constants. In Appendix~\ref{app:tunable_interaction}, we show that for all $\epsilon > 0$, terms involving $\alpha_a^0$ are irrelevant in the low energy limit.\footnote{For unscreened Coulomb interactions, integrating out the temporal sector produces an additional linear-in-$|\bs{q}|$ stiffness in the symmetric channel. The MGQCP remains stable under additional conditions derived in Appendix~\ref{app:tunable_interaction}.} Working in the Coulomb gauge, we can therefore project \eqref{eq:Dirac_contribution_alpha} to the spatially transverse sector $\alpha_i^T$. The low energy Euclidean Lagrangian for the gauge fields thus reduces to
\begin{equation}\label{eq:renormalized_alpha}
    \begin{aligned}
    L^E[\alpha_i] &\approx \frac{1}{2} \left[|\bs{q}|/g_+  + R |\bs{q}|^2 + \gamma |\omega|/|\bs{q}|\right]|\alpha_+^T|^2  \\
    &+ \frac{1}{2} \left[|\bs{q}|/g_-  + R |\bs{q}|^2 + \gamma |\omega|/|\bs{q}| \right]|\alpha_-^T|^2 \,.
    \end{aligned}
\end{equation}
At small $q$, the $|\bs{q}|$-linear terms generated by the CFT sector dominate over the diamagnetic $R |\bs{q}|^2$ term generated by Fermi surface fluctuations. Balancing the Landau damping term with the $|\bs{q}|$-linear term, we conclude that both $\alpha_+^T$ and $\alpha_-^T$ develop $z = 2$ dynamical scaling. 

The stability of generalized CF Fermi surfaces now hinges upon the competition between $\alpha^T_+$ and $\alpha^T_-$. Under the one-loop RG flow
\begin{equation*}
    \frac{d g_{\pm}}{dl} = - (g_+ + g_-) g_{\pm} \,, \quad \frac{dV_{{\rm BCS}, m}}{dl} = g_+ - g_- - V_{{\rm BCS}, m}^2 \,,
\end{equation*}
the ratio $g_+/g_-$ is invariant, while $V_{{\rm BCS}, m}$ flows towards $0$ if $g_+ > g_-$ and the microscopic value of $V_{{\rm BCS},m}^2$ is smaller than $g_+ - g_-$. To leading two orders in the $1/N_f$ expansion, the constants $g_{\pm}$ are given by~\cite{Chester2016_bootstrapQED,Giombi2016_QEDtopocurrent} 
\begin{equation}
    g_+ = 48 \pi^2 \sigma \,, \quad g_- = 1/(3\sigma) \,, 
\end{equation}
where
\begin{equation}
    \sigma(N_f) \approx \frac{1}{16} \left[1 + \frac{1}{N_f} \left(\frac{368}{9\pi^2} - 4\right) \right] \,.
\end{equation}
Plugging in $N_f = \infty$ gives $g_+/g_- \approx 5.6$ and $1/N_f$ corrections at $N_f = 6$ slightly increase the ratio to $g_+/g_- \approx 5.8$ which is significantly larger than the threshold value $1$. The pairing instability is thus averted at the critical point with a sufficiently weak microscopic BCS attraction. 

To further test the stability of Fermi surfaces, we must also rule out all potentially relevant perturbations beyond the pairing channel. First, we analyze the backreaction of $\{\alpha_i\}$ gauge fluctuations on the CFT sector. Since the CFT sector has dynamical scaling $z = 1$, the Landau damping term in $L^E_{\rm FS}[\alpha_i]$ behaves like a mass for $\alpha_{\pm}^T$. As a result, gauge fluctuations do not affect the CFT dynamics at low energy. 

A second class of potentially relevant perturbations are couplings between the Fermi surface sector and the CFT sector with a general form $\mathcal{O}_f \mathcal{O}_{\Phi}$. The most relevant gauge-invariant operator in the Fermi surface sector is the charge density $\mathcal{O}_f = \rho_{f_i}$. If the operator $\mathcal{O}_{\Phi}$ lives near a nonzero momentum, integrating out the Fermi surface generates hotspot contributions to the effective Lagrangian proportional to $\omega |\mathcal{O}_{\Phi}(\bs{q}, \omega)|^2$, which is an irrelevant perturbation to the CFT if the scaling dimension of $\mathcal{O}_{\Phi}$ is larger than $1$. This condition is satisfied by the adjoint mass operators in $N_f = 6$ QED$_3$, which are momentum-carrying operators with the lowest scaling dimensions~\cite{Lee2018_QED3QCP}. Similarly, if the operator $\mathcal{O}_{\Phi}$ lives near zero momentum, integrating out the Fermi surfaces generates a Landau damping term $(\omega/q) |\mathcal{O}_{\Phi}(\bs{q}, \omega)|^2$, which is irrelevant if the scaling dimension of $\mathcal{O}_{\Phi}$ is larger than $3/2$. In $N_f = 6$ QED$_3$, the most relevant such operator (which transforms trivially under $\alpha_i$ and lattice translation) is the singlet Dirac mass $\sum_{i=1}^2 \sum_{n=1}^3 \bar \Psi_{i,n} \Psi_{i,n}$ that tunes the system across the phase transition. By definition, the scaling dimension of $\mathcal{O}_{\Phi}$ is $3 - 1/\nu$, where $\nu$ is the correlation length exponent. As perturbative expansions predict $\nu \approx 1.5 > 2/3$~\cite{Pietro2017_QED3_eps}, the Landau damping effect is strongly irrelevant in the low energy limit. 

These arguments lead to the conclusion that the Fermi surfaces of $\{f_i\}$ survive down to zero temperature at the MGQCP. Moreover, the $\{f_i\}$ sector combines with $\alpha_{\pm}$ to form a non-Fermi liquid metal with $z = 2$ dynamical scaling, which is asymptotically decoupled from the $z = 1$ CFT sector formed by $\{\Phi_i\}$ in the low energy limit.

\section{Phenomenology of the MGQCP}

The existence of a MGQCP separating two fully gapped FCIs implies striking phenomenological signatures. We first examine thermodynamic and transport properties determined by the free energy and charge-neutral correlation functions. In thermodynamics, the heat capacity is dominated by a singular $T \log (\Lambda/T)$ contribution from the non-Fermi liquid sector, with subleading $\mathcal{O}(T^2)$ corrections from the CFT sector, analogous to the deconfined metal-insulator transitions in Refs.~\cite{Senthil2008_continuousMott,Zou2020_DMITbilayer,Zou2020_DMITU2}. The compressibility follows the Ioffe-Larkin rule $\kappa^{-1} = \kappa_f^{-1} + \kappa_{\Phi}^{-1}$. The Fermi surface contribution $\kappa_f$ approaches a constant as $T \rightarrow 0$, while the $U(1)$ continuity equation enforces the scaling relation $\kappa_{\Phi} \sim T$. Thus, the total compressibility is dominated by the CFT sector and scales as $\kappa(T) \sim T$. In DC charge transport (i.e. response to total gauge field $A_1 + A_2$), the Ioffe-Larkin rule gives $\rho_c = \rho_f + \rho_{\Phi}$, where the leading contribution to $\rho_{\Phi}$ is the universal CFT resistivity tensor $\rho^{ab}_{\Phi}(\omega, T) \approx \rho_0 f^{ab}(\omega/T)$ (with $f^{xx}(0) = 1$) and $\rho^{ab}_f(\omega, T)$ is the Fermi surface resistivity tensor which is much smaller than $\rho^{ab}_{\Phi}(\omega, T)$ in the clean limit. As a result, DC transport is dominated by the CFT contribution and $T$-dependent corrections scale as a fractional power-law of $T$ whose power is determined by the leading irrelevant operator that overlaps with the CFT current. 

Charged excitations at the MGQCP also exhibit interesting critical correlations. For both phase A and phase B, the corresponding TQFT in Eq.~\eqref{eq:fermionic_FCI_TQFT} decomposes into a sector involving only $\alpha_1$ and another sector involving only $a$, each enriched with layer charge conservation symmetries through the mutual Chern-Simons coupling between $\alpha_1, a$ and $A_1, A_2$. Let us label each quasiparticle in the gapped phase by a vector $\vec l$, where $l_1, l_2$ denote the global charge under $A_1, A_2$ and $l_3, l_4$ denote the gauge charge under $\alpha_1, a$ (due to the pairing of $f_1$ and $f_2$, the previously dynamical gauge field $\alpha_2$ is now identified with $A_1 + A_2 - \alpha_1$). Using this notation, we can write down the gapped quasiparticle excitations sourced by each matter field in the effective theory
\begin{equation}
    \begin{aligned}
    f_1^{\dagger} \sim f_2 \rightarrow (1,0,-1,0) &\,,\,\, \Phi_1^{\dagger} \sim \Phi_2 c_1^{\dagger} c_2^{\dagger} \rightarrow (0,0,1,0)  \,,\\ \Psi_1^{\dagger} \rightarrow (0,0,1,-1) &\,, \,\, \Psi_2^{\dagger} \rightarrow (1,1,-1,-1) \,, \\
    \mathcal{M}_a^{\dagger} \rightarrow (-1,-1,0,3) &\,, \,\, \psi^{\dagger} \rightarrow (0,0,0,1) \,,
    \end{aligned}
\end{equation}
where $\mathcal{M}_a^{\dagger}$ is the bare monopole operator that creates $2\pi$ flux of $a$. As the critical point is approached, $\Psi_i^{\dagger}$ and the dressed monopoles $\widetilde{\mathcal{M}}_{a,ijk}^{\dagger} \sim \mathcal{M}_a^{\dagger} \Psi_i^{\dagger} \Psi_j^{\dagger} \Psi_k^{\dagger}$ become gapless. In consequence, the composite operators 
\begin{equation*}
    \begin{aligned}
    \Phi_1^{\dagger} \sim \widetilde{\mathcal{M}}_{a,112}^{\dagger} &\,, \quad \Phi_2^{\dagger} \sim \widetilde{\mathcal{M}}_{a,221}^{\dagger} \,, \\
    (\Phi_1^{\dagger})^3 c_1 c_2 \sim \widetilde{\mathcal{M}}_{a,111}^{\dagger} &\,, \quad (\Phi_2^{\dagger})^3 c_1 c_2 \sim \widetilde{\mathcal{M}}_{a,222}^{\dagger} 
    \end{aligned}
\end{equation*}
are also gapless. Within the bosonic CFT described by \eqref{eq:bFCI_CFT}, $\Phi_1^{\dagger}$ and $\Phi_2^{\dagger}$ share a scaling dimension $x_{\Phi}$, which is distinct from the scaling dimension $x_{\Phi^3}$ of $(\Phi_i^{\dagger})^3 c_1 c_2$. Since the parton sectors $\{f_i\}$ and $\{\Phi_i\}$ decouple at low energy, the Euclidean Green's function of $c_i$ factorizes as
\begin{equation}
    G_{c_i}(\bs{r}, \tau) \approx G_{f_i}(\bs{r}, \tau) G_{\Phi_i}(\bs{r}, \tau) \sim \frac{G_{f_i}(\bs{r}, \tau)}{(|\bs{r}|^2 + \tau^2)^{x_{\Phi}}} \,, 
\end{equation}
where $x_{\Phi}$ is estimated to be $\approx 1.55$ within a large-$N_f$ expansion of QED$_3$~\cite{Pufu2013_QEDmonopole} and $G_{f_i}(\bs{r}, \tau)$ is the Green's function of the generalized CF field $f_i$
\begin{equation*}
    \begin{aligned}
    G_{f_i}(\bs{r}, \tau) &\sim_{r \rightarrow \infty} (r^{3/2} \log r)^{-1} \, \sin(k_F r - \pi/4)  \,, \\
    G_{f_i}(\bs{r}, \tau) &\sim_{\tau \rightarrow \infty} (\tau \sqrt{r})^{-1} \, \cos(k_F r - \pi/4) \,. 
    \end{aligned}
\end{equation*}
Multiplying these factors, we find a power-law decay in $G_{c_i}(\bs{r}, \tau)$ (up to logarithmic corrections). For every angle $\theta$ on the Fermi surface, we can parametrize the $f_i$ dispersion near $\theta$ as 
\begin{equation}
    \epsilon_{f,\theta}(\bs{k}) \approx v_F(\theta) k_{\perp} + \frac{1}{2} \kappa(\theta) k_{||}^2 \,,
\end{equation}
where $k_{\perp}, k_{||}$ are perpendicular/parallel components of $\bs{k}$ measured relative to $\bs{k}_F(\theta)$. The electron spectral function near $\theta$ can then be computed explicitly as
\begin{equation}
    A_{c,\theta}(\bs{k}, \omega) \sim \frac{|\omega|^{2x_{\Phi}-1}}{\log (\Lambda/\omega)} \mathcal{A}_{\theta}\left(\frac{\omega \log (\Lambda/\omega)}{v_F(\theta) k_{\perp}} \right) \,,
\end{equation}
where $\mathcal{A}_{\theta}$ is an angle-dependent scaling function. Despite the strongly damped correlations, the electron spectral function $A_{c,\theta}(\bs{k},\omega)$ displays a power-law non-analyticity at $\bs{k} = \bs{k}_F(\theta)$, indicating a sharp critical Fermi surface without fermionic quasiparticles~\cite{Senthil2008_criticalFS}. 

\subsection{Approach to criticality: zero temperature}\label{subsec:approach_zeroT}

Approaching the MGQCP from the gapped FCI-A, each matter field sources an anyon. The elementary anyons that generate everything are sourced by the matter fields
\begin{equation}
    (0,0,1,0) \sim \widetilde{\mathcal{M}}_{a,112}^{\dagger} \,, \quad (0,0,0,1) \sim \Psi_1 \, \widetilde{\mathcal{M}}_{a,112}^{\dagger} \,.
\end{equation}
Since $\widetilde{\mathcal{M}}_{a,ijk}^{\dagger}$ and $\Psi_1$ are gapless at the MGQCP, all anyons become gapless at the same critical value of the tuning parameter, but their minimum gaps vanish with different powers.

It is interesting to ask about the scaling of various energy gaps as a function of the distance $|m - m_c|$ away from the critical point. In the bosonic sector, the correlation length diverges as $\xi_{\Phi} \sim |m - m_c|^{-\nu}$ where $\nu$ is the correlation length exponent of the $N_f = 6$ QED$_3$ CFT. Using $z = 1$ scaling, one would naively conclude that the energy gaps of all anyons scale as $\Delta \sim \xi_{\Phi}^{-z} \sim |m - m_c|^{\nu}$. However, we must remember that a subset of these anyons can also be created as BdG quasiparticles or vortices of the interlayer superconductor (SC) of $\{f_i\}$. Importantly, although the boson gap scales as $\Delta_{\Phi} \sim |m-m_c|^{\nu}$, this gap is invisible to the $\{f_i, \alpha_i\}$ sector until a much lower energy scale $\Delta_f \sim \Delta_{\Phi}^2$. More precisely, recall that in any CFT, the $T = 0$ Euclidean current correlator scales as
\begin{equation}\label{eq:PiPhi_zeroTscaling}
    \Pi^{ij;ab}_{\Phi}(q, \Delta_{\Phi}) = q \, F^{ij;ab}(q/\Delta_{\Phi}) \,, 
\end{equation}
where $q = (\omega, \bs{q})$ is the three-momentum, $ij$ are layer indices, and $ab$ are spatial indices. Integrating out $\Phi_i$ thus generates a dressed Euclidean gauge field propagator in the vicinity of the MGQCP
\begin{equation*}
    G_{\alpha}^{-1}(q, \Delta_{\Phi}) \sim (R \, |\bs{q}|^2 + \gamma |\omega|/|\bs{q}|) \delta^{ab} \delta^{ij} + \Pi^{ij;ab}_{\Phi}(q, \Delta_{\Phi}) \,.
\end{equation*}
Due to the $z = 2$ scaling for $\alpha$, the momentum crossover at $q \sim \Delta_{\Phi}$ translates to an energy crossover at $\omega \sim \Delta_{\Phi}^2$. It is only for $\omega < \Delta_{\Phi}^2$ that the gauge fields detect the gap in the boson sector and begin to mediate strong attraction between $f_1$ and $f_2$. As a result, the BdG gap in the interlayer SC scales as $\Delta_f \approx c \Delta_{\Phi}^2$, where $c$ is a non-universal small prefactor. In Appendix~\ref{app:gap_scaling}, we further estimate the scaling of the vortex gap $\Delta_v$
\begin{equation}
    \Delta_v \sim \Delta_{\Phi}^{2/3} \gg \Delta_{\Phi} \gg \Delta_f \,.
\end{equation}
From this hierarchy of energy scales, we conclude that the energy gaps of all quasiparticles generated by $(1,0,-1,0)$ scale as $\Delta_f$, while the remaining quasiparticle gaps scale as $\Delta_{\Phi}$. In the special case $l_A = -1$ with minimal topological order, $(1,0,-1,0)$ maps to an electron-hole pair which is a local excitation. For more general $l_A$, $(1,0,-1,0)$ maps to a nontrivial anyon. 

\subsection{Approach to criticality: finite temperature crossover}\label{subsec:approach_finiteT}

Finally, we examine the crossover behavior of the system at fixed $m - m_c \neq 0$ as a function of temperature. Since both sides of the transition arise from pairing of generalized CFs, the crossover structures of phase A and phase B are identical and we will focus on phase A without loss of generality. At $T \gg \Delta_{\Phi}$, the Dirac fermion is effectively gapless relative to the thermal energy scale and the system behaves like the finite-temperature MGQCP at $m = m_c$. As $T$ decreases towards $\Delta_{\Phi}$, the Euclidean current-current correlation matrix of $\Phi_i$ follows the finite-$T$ generalization of \eqref{eq:PiPhi_zeroTscaling}
\begin{equation}
    \Pi^{ij;ab}_{\Phi}(q, \Delta_{\Phi}, T) = T\, F^{ij;ab}(q/T, \Delta_{\Phi}/T) \,. 
\end{equation}
Following the same argument as before, we conclude that singular gauge fluctuations only resolve the nonzero boson gap at a much lower energy scale of order $\Delta_{\Phi}^2$. For $\Delta_{\Phi}^2 \ll T \ll \Delta_{\Phi}$, the gauge fields retain $z = 2$ scaling and do not generate a pairing instability in the $\{f_i\}$ sector. The $f_i$ self energy retains a marginal Fermi liquid form $\Sigma_{f_i}(\omega) \sim \omega \log (\Lambda/\omega)$ while the $\Phi_i$ Green's function is fully gapped. As a result, the electron Green's function appears gapped and the system realizes an unpaired bilayer generalized CFL. 

\begin{figure}
    \centering
    \includegraphics[width=0.9\linewidth]{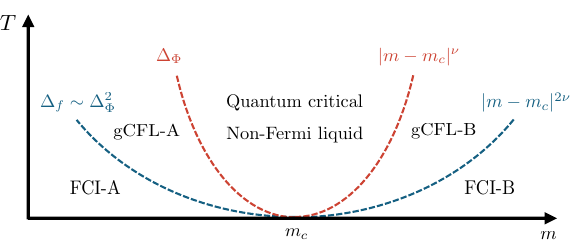}
    \caption{Finite temperature crossover near the MGQCP. An $\epsilon$-expansion estimate of the correlation length exponent $\nu \approx 1.5$ explains the shape of the crossover curves. The critical NFL and the gCFLs all have gapless Fermi surfaces. One crucial difference is that the electron is gapped in the gCFLs but gapless and strongly damped in the critical NFL.}
    \label{fig:finiteT_crossover}
\end{figure}

At a parametrically lower temperature scale $\Delta_f \approx c \Delta_{\Phi}^2$, the pairing instability occurs and the system crosses over to an ordinary gapped FCI. This crossover structure is qualitatively identical to the continuous Mott transition, with $\Delta_{\Phi}, \Delta_f$ mapping to the two energy scales $T^*, T^{**}$ in Ref.~\cite{Senthil2008_continuousMott}. A schematic of the finite-temperature phase diagram is shown in Fig.~\ref{fig:finiteT_crossover}. 

\section{Discussion}\label{sec:discussion}

Our construction of the MGQCP raises many open questions in the theory of quantum criticality. We will survey these directions and make some preliminary remarks. 

\subsection{Generalizations and controlled limits}

It is straightforward to generalize the MGQCP described by \eqref{eq:full_crit_Lagrangian} in several different directions. First, we recall that within the parton construction $\Phi_i = \chi_i \psi$, the MGQCP in \eqref{eq:full_crit_Lagrangian} arises from a jump in the Chern number of $\chi_i$ from $1$ to $-2$. An alternative possibility allowed by filling constraints is a Chern number jump from $1$ to $4$. In Appendix~\ref{app:opposite_Chernjump}, we show that this possibility also leads to an MGQCP with essentially the same phenomenology, where the bosonic sector is replaced by $N_f = 6$ QED$_3$-CS theory with Chern-Simons level $k = 6$. With other choices of periodic potentials, it is also possible to realize transitions in which $C_{\psi}$ jumps by $\pm 3$ instead. However, these transitions turn out not to be MGQCPs as we demonstrate in Appendix~\ref{app:psi_sector_transition}.

Along a different axis, we can also generalize to layer filling fractions $\nu_i = p/(2p+1)$ to obtain a family of MGQCPs indexed by $p$, which we analyze in Appendix~\ref{app:largeNf}. The bosonic sector is described by QED$_3$ with $N_f = 2|2p+1|$ and $k = 0$ or $k = N_f$, which is solvable in the $p \rightarrow \infty$ limit. This family firmly establishes the existence of metallogenic quantum criticality as a general theoretical phenomenon.

\subsection{Numerical exploration of the MGQCP}

A natural next step is to explore MGQCPs in concrete microscopic models through numerical simulations. In this work, we gave analytic arguments that a tunable family of periodic potentials \eqref{eq:periodic_potential} may realize the gapped FCIs on either side of the MGQCP. However, the argument does not rule out the possibility of a first-order transition, an intermediate phase, or a critical point described by a different universality class. A more thorough numerical investigation of this family of models will be reported elsewhere~\cite{Tan2026_bosonbilayer}. 

More generally, the power of effective field theories is that they could describe universal phenomena across a range of seemingly different microscopic models sharing the same microscopic symmetries and anomalies. Therefore, it is worth searching for other realizations of the same MGQCP in other two-component quantum Hall systems with different kinds of periodic modulations, possibly without any external magnetic field. 

\subsection{New perspective on FCI-SC transitions}

The fundamentally new phenomenon illustrated by the MGQCP is the nucleation of a positive-dimensional gapless Fermi surface (in momentum space) at a critical point between two neighboring phases without Fermi surfaces. While the neighboring phases happen to be fully gapped in our construction, this gap condition is not obviously necessary. Thinking along these lines, an intriguing question is whether a Fermi surface can emerge at an FCI-SC transition (we treat the $U(1)$ charge conservation symmetry as a global symmetry so that the SC has a gapless Goldstone mode). Numerically, recent work has shown that tuning the bandwidth/Berry curvature distribution of a $2/3$-filled Chern band can drive a direct transition from the Jain FCI to the $c_- = -1/2$ chiral topological SC~\cite{Guerci2025_SCmeltFCI,Wang2025_SCmeltFCI,guerci2026topological}. Along certain tuning trajectories, there is even preliminary evidence for a continuous transition~\cite{Wang2025_SCmeltFCI}. However, such a continuous transition appears impossible if the critical theory is described by CFTs involving gapless scalars/fermions coupled to (possibly non-Abelian) Chern-Simons theories~\cite{Shi2025_nonAbelian_TSC,Shi2026_4eTQC}. 

A radical possibility is that the transition observed in numerics realizes an MGQCP which goes beyond the framework of Chern-Simons matter CFTs. It may be fruitful to search for signatures of gapless momentum-space manifolds at the FCI-SC transition, which would provide strong evidence for the relevance of metallogenic quantum criticality. Conceptually, such a critical point would be a highly nontrivial generalization of the ideas in this work because the FCI-SC transition occurs in a single-component electronic system, where the mechanism of interlayer pairing, which is crucial to our construction, is no longer operative.

\subsection{Doping the near-critical FCIs}

Going beyond a fixed filling fraction, we can ask how proximity to the MGQCP affects the itinerant phases that are favored by doping. Recent work on doped FCIs has emphasized the important role played by the relative energy gaps of different quasiparticles in answering this question~\cite{Shi2024_doping,Shi2025_dopeMR,Shi2025_anyon_delocalization,Nosov2025_plateau,Nakajima2025_thermo_anyon}. This effect is amplified near a quantum critical point, as criticality tends to create large separations between the energy gaps of different families of quasiparticles~\cite{Divic2024_HofHubb,Pichler2025_anyonSC,Han2025_anyonexciton,Kuhlenkamp2025_HofHubb,Chen2025_HoffHubb}. Near the MGQCP, we have argued that the energy gaps of all quasiparticles generated by $(1,0,-1,0)$ are much lower than the remaining quasiparticles. In FCI-A for example, $(1,0,-1,0)$ maps to a charge-neutral anyon that carries charge $1/(l_A + 2)$ under $A_1$ and $-1/(l_A+2)$ under $A_2$. Upon introducing a small layer imbalance at fixed total charge, we therefore obtain a dilute gas of such neutral quasiparticles, with all charged quasiparticles effectively frozen. It would be interesting to explore the itinerant phases realized by this neutral gas along the lines of Ref.~\cite{Han2025_anyonexciton}.

\subsection{Emergent symmetries and classification of quantum phase transitions}

In recent years, there has been growing interest in the program of classifying gapless quantum phase transitions in terms of emergent generalized symmetries present in neighboring gapped phases~\cite{Ji2019_QCP_categorical,Chatterjee2022_symTO_I,Chatterjee2022_symTO_II,McGreevy2022_gensymreview}. From this perspective, the MGQCP poses an intriguing challenge because its critical Fermi surface respects an infinite-dimensional loop group symmetry associated with the emergent conservation of charge densities at every point on the Fermi surface~\cite{Haldane2005_FSsymmetry,Else2020_EFL,Shi2022_gifts,Kukreja2026_NFLsymmetry}, which is invisible from either neighboring phase. It is conceivable that generalizations of our construction could produce even more exotic gapless manifolds associated with distinct infinite-dimensional emergent symmetries (e.g.~\cite{Paramekanti2002_Bosemetal,Motrunich2007_Boseliquid,Lake2021_BLL}). It remains to be seen whether any such transition could be incorporated into a symmetry-based classification framework which would enrich our general understanding of quantum criticality. 
\newline 

\begin{acknowledgments}
    GPT 5.5 and 5.6 Sol were used for editorial feedback and algebraic cross-checks; all scientific claims and calculations were independently verified by the author. ZDS was supported by a Leinweber Institute for Theoretical Physics postdoctoral fellowship at Stanford University and in part by the Gordon and Betty Moore Foundation EPiQS initiative, Grant GBMF8686.01.
\end{acknowledgments}

\bibliography{MGQCP}

\appendix
\newpage
\onecolumngrid

\section{More detailed analysis of the bosonic quantum Hall bilayer}\label{app:microscopic}

In this section, we provide additional details omitted in the analysis of bosonic quantum Hall bilayers presented in Sec.~\ref{subsec:boson_bilayer}. We begin by stating the microscopic model in Appendix~\ref{subapp:microscopic_noperiodic} and describing the parton mean-field approximation that would lead to the effective gauge theory in Sec.~\ref{subsec:boson_bilayer}. An explicit quantitative determination of various mean-field parameters is complicated and deferred to future work. In Appendix~\ref{subapp:parton_gauge_theory}, we carefully derive two equivalent Lagrangian descriptions of the gauge theory for the two bosonic FCIs bFCI-A and bFCI-B introduced in Sec.~\ref{subsec:boson_bilayer}. This derivation will clarify the relation between these bosonic FCIs and the Halperin hierarchy states in Tab.~\ref{tab:topological_bFCIs}. Finally, in Appendix~\ref{subapp:parton_gauge_electronFCI}, we derive the TQFTs for the electronic FCIs obtained from pairing of generalized CFs in Sec.~\ref{subsec:boson_fermion_coupling} and comment on connections between the electronic FCIs we obtain and Halperin states in fermionic quantum Hall bilayers.

\subsection{Microscopic Hamiltonian with a tunable periodic potential}
\label{subapp:microscopic_noperiodic}

We begin with the second-quantized Hamiltonian for a bilayer bosonic system with no periodic potential
\begin{equation}
    H_0 = \sum_{i=1}^2 \int d^2 \bs{r} \, \Phi^{\dagger}_i(\bs{r}) \left[\frac{(-i\nabla - \bs{\alpha}_i)^2}{2m}\right] \Phi_i(\bs{r}) + \frac{1}{2} \int d^2 \bs{r}\,d^2 \bs{r}'\, \delta\rho(\bs{r}) \frac{V_0}{|\bs{r}-\bs{r}'|} \delta\rho(\bs{r}') \,,
\end{equation}
where $\delta\rho \equiv \rho_{\Phi_1} + \rho_{\Phi_2} - \bar\rho$ includes a uniform neutralizing background and the Coulomb energy scale $V \equiv V_0/l_{\rm mag}$. We choose the magnetic field $\nabla \times \bs{\alpha}_1 = \nabla \times \bs{\alpha}_i = B$ and boson densities $\rho_{\Phi_1} = \rho_{\Phi_2}$ such that each boson is at Landau level filling $\nu_i = 1/3$ and the cyclotron gap $\omega_c \equiv B/m_{\Phi}$ is much larger than the interaction scale $V$. 

The crucial new ingredient is a square lattice periodic potential $W(\bs{r})$. Choosing units in which the unit cell area is $a^2 = 1$ and the magnetic flux through each unit cell is $4\pi$, we can write the full Hamiltonian as
\begin{equation}
    W(\bs{r}) = W \cos (2\pi x) \cos (2\pi y) + \alpha \, W \left[\cos(2\pi x)+\cos(2\pi y)\right]  \,, \quad H = H_0 + \sum_{i=1}^2 \int d^2 \bs{r} \, \Phi_i^{\dagger}(\bs{r}) \Phi_i(\bs{r}) \, W(\bs{r}) \,.
\end{equation}
The boson density per square lattice unit cell is then $\nu^{lat}_i = 2/3$, which is twice as large as the Landau level filling in the absence of the periodic potential. 

To describe a fractional quantum Hall state, we will follow the main text and use the fermionic parton decomposition $\Phi_i =\chi_i \psi$. This decomposition introduces a $U(1)$ gauge redundancy which can be removed by coupling $\chi_i$ and $\psi$ to a dynamical $U(1)$ gauge field $a$. Without loss of generality, we choose $\chi_i$ to carry charge $-1$ and $\psi$ to carry charge $+1$ under the emergent gauge field $a$. Interesting correlated phases for the original bosons $\Phi_i$ can be obtained from free-fermion phases of $\chi_i$ and $\psi$. This strategy motivates a general class of parton-based variational wavefunctions
\begin{equation}\label{eq:projected_wavefunction}
    \ket{\Psi}_{\Phi} = \mathcal{P}_G \left[\ket{\Psi}_{\chi} \otimes \ket{\Psi}_{\psi}\right] \,, 
\end{equation}
where $\ket{\Psi}_{\chi}, \ket{\Psi}_{\psi}$ are free-fermion wavefunctions with some variational parameters and $\mathcal{P}_G$ is a projection onto the subspace satisfying the local constraint $\rho_{\Phi_i}(\bs{r}) = \rho_{\chi_i}(\bs{r}) = \rho_{\psi}(\bs{r})/2$. The projection guarantees that $\ket{\Psi}_{\Phi}$ defines a wavefunction in the correct microscopic Hilbert space. 

To motivate these wavefunctions analytically, one can postulate a mean-field ansatz in which the emergent magnetic field $\nabla \times \bs{a}$ takes on a static nonzero value on average. The outcome of such an approximate treatment is a general parton Hamiltonian of the form
\begin{equation}
    H = \sum_{i=1}^2 H[\chi_i, \alpha_i - \bar a - \delta a] + H[\psi, \bar a + \delta a] + H_{\rm int} \,,
\end{equation}
where 
\begin{equation}
    \begin{aligned}
    H[\chi_i, \alpha_i - \bar a - \delta a] &= \int d^2 \bs{r} \chi_i^{\dagger}(\bs{r}) \varepsilon_{\chi}(-i\nabla - \bs{\alpha}_i + \bs{\bar a} + \bs{ \delta a}) \chi_i(\bs{r}) + \int d^2 \bs{r} \, \left[\alpha_1^0(\bs{r}) - a^0(\bs{r}) - \delta a^0(\bs{r}) \right] \rho_{\chi_i}(\bs{r}) \,, \\
    H[\psi, a] &= \int d^2 \bs{r} \psi^{\dagger}(\bs{r}) \varepsilon_{\psi}(-i\nabla - \bs{\bar a} - \bs{\delta a}) \psi(\bs{r}) + \int d^2 \bs{r} \, [\bar a^0(\bs{r}) + \delta a^0(\bs{r})] \rho_{\psi}(\bs{r})  \,,
    \end{aligned}
\end{equation}
and $H_{\rm int}$ encodes interactions effects beyond the mean-field approximation. From this Hamiltonian, one can infer the Lagrangian, which we presented in the main text using the compact notation
\begin{equation}
    L[\{\Phi_i\}, \{\alpha_i\}] = \sum_{i=1}^2 L[\chi_i, \alpha_i - a] + L[\psi, a] \,. 
\end{equation}
Setting $\delta a = 0$ freezes the gauge fluctuations and defines the parton mean-field approximation. Including $\delta a$ fluctuations recovers the full low energy theory which provides a faithful field-theoretic description of the projected wavefunction in \eqref{eq:projected_wavefunction}. Note that the dispersion relations $\varepsilon_{\chi}, \varepsilon_{\psi}$ must be determined self-consistently by including renormalization effects from the Coulomb interactions within some approximation scheme. 

A detailed solution of the parton mean-field self-consistency equations is beyond the scope of this work (this problem is especially challenging when microscopic interactions are projected onto the lowest Landau level, which is the appropriate setup for our model). In what follows, we will instead postulate particularly simple mean-field states that are consistent with existing unbiased numerical simulations in the $W = 0$ region of the phase diagram. A more thorough numerical investigation of the general validity of our parton ansatz will be reported elsewhere~\cite{Tan2026_bosonbilayer}. 

\subsection{Derivation of topological quantum field theories for bosonic FCIs}\label{subapp:parton_gauge_theory}

We begin with the limit $W = 0$, where the parton mean-field approximation reduces to the composite fermion mean-field approximation familiar in conventional quantum Hall bilayers~\cite{Geraedts2017_onethird_Bbilayer}. Recall that each $\Phi_i$ is at Landau level filling $\nu_i = 1/3$ relative to the external magnetic field $\nabla \times \bs{\alpha}_i = B$. The preferred static flux configuration is $\nabla \times \bs{a} = \frac{2}{3} \nabla \times \bs{\alpha}_i$. Under this flux configuration, each fermionic parton $\chi_i, \psi$ completely fills a single Landau level. As a result, up to Gaussian prefactors, a prototypical wavefunction is a product of three integer quantum Hall wavefunctions
\begin{equation}\label{eq:wavefunction_221}
    \Psi_{\rm (221)}(\{z_1^a, z_2^b\}) = \Psi_{\rm IQH}(\{z_1^a\}) \Psi_{\rm IQH}(\{z_2^a\}) \Psi_{\rm IQH}(\{z_1^a, z_2^b\}) \sim \prod_{a<b}(z_1^a-z_1^b)^2 \prod_{a<b}(z_2^a-z_2^b)^2 \prod_{a,b}(z_1^a-z_2^b) \,.
\end{equation}
We recognize this wavefunction as the Halperin (221) state. The topological order of this wavefunction is straightforward to derive. Integrating out the gapped partons in the Lagrangian $\sum_{i=1}^2 L[\chi_i, \alpha_i - a] + L[\psi, a]$ gives the low energy effective field theory
\begin{equation}\label{eq:parton_bFCIA_step1}
    L_{\rm bFCI}^{(A)} = \mathrm{CS}[\alpha_1 - a, g] + \mathrm{CS}[\alpha_2 - a, g] + \mathrm{CS}[a,g] \,, \quad \mathrm{CS}[a,g] \equiv \frac{1}{4\pi} a da + 2 \mathrm{CS}_g \,,
\end{equation}
where $\mathrm{CS}[a,g]$ is the response theory of a fermionic integer quantum Hall state with Hall conductance $\sigma_{xy} = 1$ to background electromagnetic field $a$ and background metric $g$. The combination of the Chern-Simons term for the $\mathrm{spin}_{\mathbb{C}}$ connection $a$ and the gravitational Chern-Simons term for $g$ guarantees that $\mathrm{CS}[a,g]$ is well-defined for arbitrary oriented spacetime manifolds. By expanding every term explicitly, we recover the TQFT claimed in the main text
\begin{equation}
    L_{\rm bFCI}^{(A)} = \mathrm{CS}[a,g] - \frac{1}{2\pi} a d (\alpha_1 + \alpha_2) + \sum_{i=1}^2 \frac{1}{4\pi} \alpha_i d \alpha_i \,. 
\end{equation}
It is useful to obtain an alternative representation of this TQFT to make a more direct connection with the parton wavefunction \eqref{eq:wavefunction_221}. We can introduce auxiliary $U(1)$ gauge fields $\beta_1, \beta_2$ to rewrite \eqref{eq:parton_bFCIA_step1} as 
\begin{equation}
    L_{\rm bFCI}^{(A)} = - \frac{1}{4\pi} \beta_1 d \beta_1 + \frac{1}{2\pi} \beta_1 d (\alpha_1 - a) - \frac{1}{4\pi} \beta_2 d \beta_2 + \frac{1}{2\pi} \beta_2 d (\alpha_2 - a) + \mathrm{CS}[a,g] \,. 
\end{equation}
Integrating out $a$ gives a simple $K$-matrix TQFT
\begin{equation}
    L_{\rm bFCI}^{(A)} = - \frac{1}{4\pi} \beta^T \, K_A \, d \beta + \sum_{i=1}^2 \frac{1}{2\pi} \beta_i d \alpha_i \,, \quad K_A = \begin{pmatrix}
        2 & 1 \\ 1 & 2 
    \end{pmatrix} \,. 
\end{equation}
This final form is a standard K-matrix representation of the Halperin (221) topological order. The entries in the K-matrix have a direct correspondence with the powers of $(z_i^a - z_j^b)$ in the wavefunction \eqref{eq:wavefunction_221}. 

It is straightforward to verify that this effective theory has Hall conductance $\sigma_{xy} = 2/3$, counterflow Hall conductance $\sigma^{\rm cf}_{xy} = 2$, chiral central charge $c_- = 2$, and three Abelian anyons. In fact, this is the unique topological order with $U(1)_1 \times U(1)_2$ symmetry that satisfies these universal properties~\cite{Cheng2025_orderingFQH}. Numerical simulations confirm that this is indeed the ground state of the bosonic quantum Hall bilayer for a range of $d < l_{\rm mag}$ (including at $d = 0$) in the absence of any periodic potential~\cite{Geraedts2017_onethird_Bbilayer}. Thus, our parton mean-field approximation is a sensible starting point for an analytic understanding of fractionalization in the bosonic quantum Hall bilayer. 

Upon turning on a nonzero periodic potential $W$, we assume that the total static gauge flux $\nabla \times \bs{a}$ per unit cell remains fixed, while the renormalized band dispersions $\varepsilon_{\chi}, \varepsilon_{\psi}$ vary continuously. Recall that the periodic potential is chosen so that each unit cell encloses $4\pi$ flux. In units where the lattice unit cell has area $1$, the average static flux in each unit cell thus evaluate to
\begin{equation}
    \int_{\rm u.c.} \frac{\nabla\times \bs{a}}{2\pi}= \frac{4}{3} \,, \quad
    \int_{\rm u.c.} \frac{\nabla\times (\bs{A}_i - \bs{a})}{2\pi}= \frac{2}{3} \,.
\end{equation}
With this flux assignment, the $\chi_i, \psi$ partons can form gapped Chern insulators with Chern numbers $C_{\chi_i} = C_{\psi} = 1$ mod 3. In the small $W$ regime, these Chern insulators are smoothly connected to the continuum $\sigma_{xy} = 1$ integer quantum Hall states formed by $\chi_i, \psi$ in the $W = 0$ limit. The resulting topological order is identical to the Halperin (221) state previously constructed. 

As $W$ increases, it is possible that $\chi_i$ or $\psi$ undergoes a band inversion transition. Due to arguments in Sec.~\ref{subsec:boson_bilayer}, the simplest possibility compatible with symmetries and anomalies of the microscopic model is that $\chi_i$ undergoes a band-inversion transition across which its Chern number jumps from $1$ to $-2$. Two alternative possibilities are also considered in other Appendices. In particular, if $C_{\chi_i}$ jumps from $1$ to $4$ instead, we obtain another MGQCP with qualitatively similar properties (see Appendix~\ref{app:opposite_Chernjump}); if $C_{\psi}$ jumps away from $1$, the MGQCP is avoided and replaced by an ordinary CFT (see Appendix~\ref{app:psi_sector_transition}). 

Focusing now on the simplest possibility in which $C_{\chi_i}$ jumps from $1$ to $-2$, we can integrate out all the fermionic partons and recover the TQFT for bFCI-B stated in Sec.~\ref{subsec:boson_bilayer} of the main text
\begin{equation}\label{eq:parton_bFCIB_step1}
    L_{\rm bFCI}^{(B)} = -2 \mathrm{CS}[\alpha_1 - a, g] - 2 \mathrm{CS}[\alpha_2-a, g] + \mathrm{CS}[a,g] = - 3 \mathrm{CS}[a,g] + \frac{2}{2\pi} a d (\alpha_1 + \alpha_2) - \sum_{i=1}^2 \frac{2}{4\pi} \alpha_i \, d \alpha_i \,. 
\end{equation}
The associated parton wavefunction takes the form
\begin{equation}
    \begin{aligned}
    \Psi_{\rm bFCI-B}(\{z^a_1, z^b_2\}) &= \Psi_{C = -2}(\{z_1^a\}) \Psi_{C = -2}(\{z_2^a\}) \Psi_{C = 1}(\{z_1^a, z_2^b\}) \,,
    \end{aligned}
\end{equation}
where $\Psi_C$ represents a wavefunction of a Chern insulator with Chern number $C$. Note that due to the presence of a periodic potential, $\Psi_C$ is not equivalent to the wavefunction of $C$ filled Landau levels. 

It is again helpful to recast the effective field theory in a K-matrix form. Towards that end, we again introduce a pair of auxiliary $U(1)$ gauge fields $\beta_1, \beta_2$ to rewrite \eqref{eq:parton_bFCIB_step1} as
\begin{equation}
    \begin{aligned}
        L_{\rm bFCI}^{(B)} &= \sum_{i=1}^2 \left[- \mathrm{CS}[\alpha_i - a, g] + \frac{1}{4\pi} \beta_i d \beta_i + \frac{1}{2\pi} \beta_i d (\alpha_i - a)\right] + \mathrm{CS}[a,g] \\
        &= \sum_{i=1}^2 \left[-\frac{1}{4\pi} \alpha_i d \alpha_i + \frac{1}{4\pi} \beta_i d \beta_i + \frac{1}{2\pi} \beta_i d \alpha_i\right] - \mathrm{CS}[a,g] + \frac{1}{2\pi} a d (\alpha_1 + \alpha_2 - \beta_1 - \beta_2) \,. 
    \end{aligned}
\end{equation}
Shifting $a \rightarrow a - \beta_1 - \beta_2$ gives
\begin{equation}
    \begin{aligned}
        L_{\rm bFCI}^{(B)} &= \sum_{i=1}^2 \left[-\frac{1}{4\pi} \alpha_i d \alpha_i + \frac{1}{4\pi} \beta_i d \beta_i + \frac{1}{2\pi} \beta_i d \alpha_i\right] - \mathrm{CS}[a,g] + \frac{1}{2\pi} a d (\alpha_1 + \alpha_2 - \beta_1 - \beta_2) \\
        &- \frac{1}{4\pi}(\beta_1 + \beta_2) d (\beta_1 + \beta_2) + \frac{1}{2\pi} a d (\beta_1 + \beta_2) - \frac{1}{2\pi} (\beta_1 + \beta_2) d (\alpha_1 + \alpha_2 - \beta_1 - \beta_2) \\
        &= \frac{1}{4\pi} (\beta_1 + \beta_2) d (\beta_1 + \beta_2) + \sum_{i=1}^2 \left[\frac{1}{4\pi} \beta_i d \beta_i - \frac{1}{4\pi} \alpha_i d \alpha_i\right] - \mathrm{CS}[a,g] + \frac{1}{2\pi} a d (\alpha_1 + \alpha_2) - \frac{1}{2\pi} \beta_1 d \alpha_2 - \frac{1}{2\pi} \beta_2 d \alpha_1 \,.
    \end{aligned}
\end{equation}
Integrating out the $\mathrm{spin}_{\mathbb{C}}$ connection $a$ and relabeling $\beta_1 \leftrightarrow \beta_2$, we find a familiar form
\begin{equation}
    \begin{aligned}
    L_{\rm bFCI}^{(B)} &= \frac{1}{4\pi} (\beta_1 + \beta_2) d (\beta_1 + \beta_2) + \frac{1}{4\pi} \beta_1 d \beta_1 + \frac{1}{4\pi} \beta_2 d \beta_2 - \frac{1}{2\pi} \beta_1 d \alpha_1 - \frac{1}{2\pi} \beta_2 d \alpha_2 + \frac{1}{2\pi} \alpha_1 d \alpha_2 \\
    &= - \frac{1}{4\pi} \beta^T K_B \, d \beta - \sum_{i=1}^2 \frac{1}{2\pi} \beta_i d \alpha_i + \frac{1}{2\pi} \alpha_1 d \alpha_2 \,,
    \end{aligned}
\end{equation}
where $K_B = - K_A$. We recognize the final line as a boson integer quantum Hall state stacked with the time-reversal of the Halperin (221) state. This TQFT has chiral central charge $c_- = -2$, Hall conductance $\sigma_{xy} = 4/3$, and counterflow Hall conductance $\sigma^{\rm cf}_{xy} = -4$. Remarkably, the universal topological properties of this state are identical to the bilayer bosonic Jain state at a different Landau level filling $\nu_i = 2/3$ \textit{without any external periodic potential}, which has been observed in DMRG studies of bosonic quantum Hall bilayers~\cite{Geraedts2017_onethird_Bbilayer}. This calculation justifies the identification of topological orders in Tab.~\ref{tab:topological_bFCIs}. 

\subsection{Derivation of topological quantum field theories for fermionic FCIs}\label{subapp:parton_gauge_electronFCI}

Finally, we combine the bosonic TQFTs with the fermionic sectors and derive the fermionic TQFTs in \eqref{eq:fermionic_FCI_TQFT}. 

We begin with the A-side of the phase transition. For a fermionic bilayer system forming a gapped interlayer superconductor with angular momentum $l_A$, the general TQFT is 
\begin{equation}
    L_{\rm SC}^{(l_A)}[f_i, A_i - \alpha_i] = l_A \mathrm{CS}[A_1 - \alpha_1, g] + \frac{1}{2\pi} \beta d (A_1 + A_2 - \alpha_1 - \alpha_2) \,. 
\end{equation}
Combining this TQFT with the bosonic sector, we find 
\begin{equation}
    L_{\rm FCI-A} = l_A \mathrm{CS}[A_1 - \alpha_1, g] + \frac{1}{2\pi} \beta d (A_1 + A_2 - \alpha_1 - \alpha_2) + \frac{1}{4\pi} \alpha_1 d \alpha_1 + \frac{1}{4\pi} \alpha_2 d \alpha_2 + 3 \mathrm{CS}[a,g] - \frac{1}{2\pi} a d (\alpha_1 + \alpha_2) \,. 
\end{equation}
Integrating out $\beta$ identifies $\alpha_2$ with $A_1 + A_2 - \alpha_1$. Further integrating out $\alpha_2$ gives
\begin{equation}
    L_{\rm FCI-A} = l_A \mathrm{CS}[A_1 - \alpha_1, g] + \frac{1}{4\pi} \alpha_1 d \alpha_1 + \frac{1}{4\pi} (A_1 + A_2 - \alpha_1) d (A_1 + A_2 - \alpha_1) + 3 \mathrm{CS}[a,g] - \frac{1}{2\pi} a d (A_1 + A_2) \,,
\end{equation}
which reproduces the first line of \eqref{eq:fermionic_FCI_TQFT}. Importantly, $\alpha_1$ is a $U(1)$ gauge field while $a$ is a $\mathrm{spin}_{\mathbb{C}}$ connection. This theory describes a gapped topological order except for $l_A = -2$, where the relative $U(1)$ symmetry is spontaneously broken. The resulting state is an exciton SF* coexisting with an Abelian topological order described by the $a$-sector. For all $l_A \neq -2$, it is straightforward to compute the following quantized invariants
\begin{equation}
    \sigma_{xy} = \frac{2}{3} \,, \quad \sigma^{\rm cf}_{xy} = \frac{2l_A}{l_A+2} \,, \quad c_- = 2 + l_A - \sgn(2+l_A) \,, 
\end{equation}
where $\sigma_{xy}$ is the charge response to $(A_1 + A_2)/2$ and $\sigma^{\rm cf}_{xy}$ is the counterflow response to $(A_1 - A_2)/2$. Note that there is no mixing between these channels due to layer exchange symmetry. These values reproduce the first row of Tab.~\ref{tab:gapped_FCIs} in the main text. 

The special case $l_A = -1$ is particularly interesting. In this special case, it is helpful to rewrite FCI-A by introducing two auxiliary $U(1)$ gauge fields $\beta_1, \beta_2$
\begin{equation}
    L_{\rm FCI-A} = - \mathrm{CS}[A_1 - \alpha_1, g] + \frac{1}{2\pi} \beta d (A_1 + A_2 - \alpha_1 - \alpha_2) + \mathrm{CS}[a,g] - \frac{1}{4\pi} \beta_1 d \beta_1 + \frac{1}{2\pi} \beta_1 d (\alpha_1 - a) - \frac{1}{4\pi} \beta_2 d \beta_2 + \frac{1}{2\pi} \beta_2 d (\alpha_2 - a) \,. 
\end{equation}
Now let us first integrate out $a$ and then integrate out $\beta$ to get
\begin{equation}
    L_{\rm FCI-A} = - \mathrm{CS}[A_1 - \alpha_1, g] - \frac{2}{4\pi} \beta_1 d \beta_1 - \frac{2}{4\pi} \beta_2 d \beta_2 - \frac{1}{2\pi} \beta_1 d \beta_2 + \frac{1}{2\pi} \beta_1 d \alpha_1 + \frac{1}{2\pi} \beta_2 d (A_1 + A_2 - \alpha_1) \,. 
\end{equation}
Finally integrating out $\alpha_1$ gives
\begin{equation}
    \begin{aligned}
    L_{\rm FCI-A} &= - \mathrm{CS}[A_1,g] - \frac{2}{4\pi} \beta_1 d \beta_1 - \frac{2}{4\pi} \beta_2 d \beta_2 - \frac{1}{2\pi} \beta_1 d \beta_2 + \frac{1}{2\pi} \beta_2 d (A_1 + A_2) + \mathrm{CS}[\beta_1 - \beta_2 + A_1, g] \\
    &= - \frac{1}{4\pi} \beta^T \, \tilde K_A \, d \beta + \frac{1}{2\pi} \beta_1 d A_1 + \frac{1}{2\pi} \beta_2 d A_2 \,,
    \end{aligned}
\end{equation}
where we defined a two-component fermionic K-matrix
\begin{equation}
    \tilde K_A = \begin{pmatrix}
        1 & 2 \\ 2 & 1 
    \end{pmatrix} \,.
\end{equation}
This two-component K-matrix defines the Halperin (112) state, which is precisely the topological order observed in numerical simulations of fermionic quantum Hall bilayers at layer Landau level filling 1/3 and small interlayer distance~\cite{Geraedts2015_onethird_Fbilayer}. 

The exact same exercise can be done on the B-side of the phase transition. Assuming a pairing angular momentum $l_B$, the effective theory for the fermionic FCI is
\begin{equation}
    \begin{aligned}
    L_{\rm FCI-B} = \, &l_B \mathrm{CS}[A_1 - \alpha_1, g] + \frac{1}{2\pi} \beta d (A_1 + A_2 - \alpha_1 - \alpha_2) \\
    &- 3\mathrm{CS}[a,g] + \frac{2}{2\pi} a d (\alpha_1 + \alpha_2) - \frac{2}{4\pi} \alpha_1 d \alpha_1 - \frac{2}{4\pi} (A_1 + A_2 - \alpha_1) d (A_1 + A_2 - \alpha_1) \,.
    \end{aligned}
\end{equation}
Integrating out $\beta$ and $\alpha_2$ reproduces the second line of \eqref{eq:fermionic_FCI_TQFT}
\begin{equation}
    L_{\rm FCI-B} = l_B \mathrm{CS}[A_1 - \alpha_1, g] - \frac{2}{4\pi} \alpha_1 d \alpha_1 - \frac{2}{4\pi} (A_1 + A_2 - \alpha_1) d (A_1 + A_2 - \alpha_1) - 3\mathrm{CS}[a,g] + \frac{2}{2\pi} a d (A_1 + A_2)  \,.
\end{equation}
This theory describes a gapped topological order except for $l_B = 4$, where integrating $\alpha_1$ Higgses the combination $2A_1 - 2A_2$. The resulting phase is a condensate of doubled excitons coexisting with an Abelian topological order described by the $a$ sector. For $l_B \neq 4$, it is straightforward to compute the following quantized invariants
\begin{equation}
    \sigma_{xy} = 4/3 \,, \quad \sigma^{\rm cf}_{xy} = \frac{4l_B}{4-l_B} \,, \quad c_- = -2 + l_B + \sgn(4 - l_B) \,.
\end{equation}
These values reproduce the second row of Tab.~\ref{tab:gapped_FCIs} in the main text. 

For general $l_B \neq 4$, the topological order FCI-B has $(l_B-4)3$ anyons. The choice of $l_B$ that produces the minimal topological order is thus $l_B = 3$ or $l_B = 5$. These two cases have the same chiral central charge $c_- = 2$ but differ in the counterflow Hall conductivity. 

Finally, let us make an intriguing observation about the relationship between FCI-A and FCI-B. For generic $l_A/l_B$, there is no simple transformation that relates them. However, if we turn on a weak tunneling between the two layers, the counterflow Hall conductivity ceases to be well-defined and all quantized invariants of $l_A=-1$ match with $l_A = -3$. A similar equivalence holds for $l_B = 3$ and $l_B = 5$. Since $\sigma_{xy}$ and $c_-$ uniquely specify the minimal topological order~\cite{Cheng2025_orderingFQH} in the presence of total charge $U(1)$ symmetry, we can identify $l_A=-1, -3$ as a single symmetry-enriched topological order $\mathrm{FCI-A}_{\rm min}$ and $l_B=3,5$ as a single symmetry-enriched topological order $\mathrm{FCI-B}_{\rm min}$. Remarkably, observe that
\begin{equation}
    \sigma_{xy} = \begin{cases}
        2/3 & \mathrm{FCI-A}_{\rm min} \\
        4/3 & \mathrm{FCI-B}_{\rm min}
    \end{cases} \,, \quad c_- = \begin{cases}
        0 & \mathrm{FCI-A}_{\rm min} \\
        2 & \mathrm{FCI-B}_{\rm min}
    \end{cases} \,.
\end{equation}
Therefore, $\mathrm{FCI-A}_{\rm min}$ and $\mathrm{FCI-B}_{\rm min}$ are precisely related by particle-hole conjugation in the fermionic bilayer. We emphasize that this simple relation relies crucially on the weak breaking of layer pseudospin symmetry. If the symmetry was not broken, additional stacking by invertible phases is needed to account for the difference in counterflow Hall conductivity between the A side and the B side. 

\section{Effective gauge field Lagrangian with tunable microscopic interaction}\label{app:tunable_interaction}

In this section, we analyze the effective Lagrangian for the gauge fields $\{\alpha_i\}$ when the bosonic partons $\{\Phi_i\}$ are tuned to the CFT point and the fermionic partons $\{f_i\}$ form interacting Fermi liquids. The structure of the gauge field Lagrangian will depend on the long-distance asymptotics of microscopic density-density interactions. For maximal generality, we will consider an intralayer interaction $V(\bs{r})$ and an interlayer interaction $U(\bs{r})$ with tunable power-law decay
\begin{equation}
    V(\bs{r}) = \frac{V_0}{|\bs{r}|^{1+\epsilon}} \,, \quad U(\bs{r}) = \frac{V_0}{(\sqrt{d^2 + |\bs{r}|^2})^{1+\epsilon}} \,, \quad V(\bs{q}) \approx A_{\epsilon} \, |\bs{q}|^{\epsilon - 1} \,, \quad U(\bs{q}) \approx B_{\epsilon} \, \left(\frac{|\bs{q}|}{2d}\right)^{(\epsilon-1)/2}\, K_{\frac{1-\epsilon}{2}}(|\bs{q}| d) \,,
\end{equation}
where $K_{\mu}$ is the modified Bessel function of the second kind and the positive constants $A_{\epsilon}, B_{\epsilon}$ are 
\begin{equation}
    A_{\epsilon} = 2^{1-\epsilon} \pi V_0 \frac{\Gamma \left(\frac{1-\epsilon}{2}\right)}{\Gamma \left(\frac{1+\epsilon}{2}\right)} \,, \quad B_{\epsilon} = \frac{2\pi V_0}{\Gamma \left(\frac{1+\epsilon}{2}\right)} \,. 
\end{equation}
The corresponding critical Euclidean Lagrangian takes the form
\begin{equation}
    \begin{aligned}
    L &= \sum_{i=1}^2 L[f_i, A_i - \alpha_i] + \sum_{i=1}^2 \sum_{n=1}^3 L[\Psi_{i,n}, \alpha_i - a] + \frac{i}{4\pi} a d (\alpha_1 + \alpha_2) - \frac{i}{8\pi} \alpha_1 d \alpha_1 - \frac{i}{8\pi} \alpha_2 d \alpha_2 \\
    &+ \frac{1}{2} \left[|\rho_{f_1}(\bs{q, \omega})|^2 + |\rho_{f_2}(\bs{q}, \omega)|^2\right] V(\bs{q}) + \rho_{f_1}(\bs{q}, \omega) \rho_{f_2}(\bs{q}, \omega) U(\bs{q})  \,. 
    \end{aligned}
\end{equation}
In the spatially transverse gauge $\nabla \cdot \bs{\alpha}_{\pm} = 0$, integrating out Fermi surfaces $\{f_i\}$ at the RPA level gives 
\begin{equation}
    \begin{aligned}
    L_{\rm FS}[\alpha_i] &= \frac{1}{2} \frac{\Pi(\bs{q}, \omega)}{1 + \left[V(\bs{q}) + U(\bs{q})\right] \Pi(\bs{q}, \omega)} |\alpha_+^0|^2 + \frac{1}{2} \Pi_T(\bs{q}, \omega) |\alpha_+^T|^2\\
    &+ \frac{1}{2} \frac{\Pi(\bs{q}, \omega)}{1 + \left[V(\bs{q}) - U(\bs{q})\right] \Pi(\bs{q}, \omega)} |\alpha_-^0|^2 + \frac{1}{2} \Pi_T(\bs{q}, \omega) |\alpha_-^T|^2\,,
    \end{aligned}
\end{equation}
where $\Pi, \Pi_T$ are the Coulomb-irreducible density-density and current-current correlation functions and we have dropped the $(\bs{q}, \omega)$ label for each gauge field for notational convenience. We assume inversion symmetry so that the mixed density-current correlators vanish identically. 

In a Fermi liquid, $\Pi$ and $\Pi_T$ (in Euclidean signature) take the general form
\begin{equation}
    \Pi(\bs{q}, \omega) \approx \chi_0 \left[1 - \mathcal{O}\left(\frac{\omega}{q}\right)\right] \,, \quad \Pi_T(\bs{q}, \omega) \approx C \, |\bs{q}|^2 + \frac{\gamma |\omega|}{|\bs{q}|} \,. 
\end{equation}
Plugging this general form into $L_{\rm FS}$, we see that 
\begin{equation}\label{eq:LFS_general_eps}
    \begin{aligned}
        L_{\rm FS}[\alpha_i] &\approx \frac{1}{2} \,\frac{\chi_0}{1 + \left[V(\bs{q}) + U(\bs{q})\right] \chi_0}\,|\alpha_+^0|^2 + \frac{1}{2} \left[C \, |\bs{q}|^2 + \frac{\gamma |\omega|}{|\bs{q}|} \right] \, |\alpha_+^T|^2 \\
        &+ \frac{1}{2} \chi_0 \, |\alpha_-^0|^2 + \frac{1}{2} \left[C \, |\bs{q}|^2 + \frac{\gamma |\omega|}{|\bs{q}|} \right] \, |\alpha_-^T|^2 \,.
    \end{aligned}
\end{equation}
On the other hand, the CFT sector contributes an effective kinetic term for $\alpha_i$ 
\begin{equation}
    L_{\rm crit, \Phi}[\alpha_i] = \frac{1}{2} \alpha_+^{\mu} \left(\frac{\sqrt{\omega^2 + |\bs{q}|^2}}{48 \pi^2 \sigma} P_{\mu\nu}\right) \alpha_+^{\nu} + \frac{i}{8\pi} \alpha_+ d \alpha_+ + \frac{1}{2} \alpha_-^{\mu} (3 \sigma \sqrt{\omega^2 + |\bs{q}|^2} P_{\mu\nu}) \alpha_-^{\nu} - \frac{i}{8\pi} \alpha_- d \alpha_- \,. 
\end{equation}
Working in the transverse gauge and taking the limit $\omega \ll q$ with Fourier convention $\nabla_j = i q_j$, the Lagrangian simplifies to
\begin{equation}\label{eq:LPhi_general_eps}
    \begin{aligned}
    L_{\rm crit, \Phi}[\alpha_i] &= \frac{1}{2} \left(\frac{|\bs{q}|}{48\pi^2 \sigma}\right) \left[|\alpha_+^0|^2 + |\alpha_+^T|^2\right] - \frac{|\bs{q}|}{8\pi} \left[(\alpha_+^0)^* \, \alpha_+^T - (\alpha_+^T)^* \, \alpha_+^0\right] \\
    &+ \frac{1}{2} (3 \sigma |\bs{q}|) \left[|\alpha_-^0|^2 + |\alpha_-^T|^2\right] + \frac{|\bs{q}|}{8\pi} \left[(\alpha_-^0)^* \, \alpha_-^T - (\alpha_-^T)^* \, \alpha_-^0\right] \,. 
    \end{aligned}
\end{equation}

We now combine the Fermi surface contribution $L_{\rm FS}[\alpha_i]$ with the CFT contribution $L_{\rm crit, \Phi}[\alpha_i]$. The properties of the total effective Lagrangian depend on the choice of $\epsilon$. We will analyze two qualitatively different cases in turn: (1) $\epsilon > 0$, (2) $\epsilon = 0$ (unscreened Coulomb).

\subsection{Screened Coulomb interaction: $\epsilon > 0$}

When $\epsilon > 0$, the dominant contribution to the effective Lagrangian of $\alpha_{\pm}^0$ comes from the Fermi surface term \eqref{eq:LFS_general_eps}. $\alpha_-^0$ is completely screened by the density fluctuations of $\{f_i\}$ and integrating out $\alpha_-^0$ only gives analytic corrections to the $\alpha_-^T$ kinetic term. On the other hand, integrating out $\alpha_+^0$ generates a kinetic term for $\alpha_+^T$ which scales as $|\bs{q}|^{1 + \epsilon}$. Clearly, this kinetic term is subdominant relative to the leading $|\bs{q}|$-linear contribution from the CFT sector. Putting together these observations, we conclude that the low energy Lagrangian simplifies to
\begin{equation}
    L_{\rm eff}[\alpha_i] \approx \frac{1}{2} \left[\frac{|\bs{q}|}{g_+} + \frac{\gamma |\omega|}{|\bs{q}|}\right] \, |\alpha_+^T|^2 + \frac{1}{2} \left[\frac{|\bs{q}|}{g_-} + \frac{\gamma |\omega|}{|\bs{q}|}\right] \, |\alpha_-^T|^2 \,, \quad g_+ = 4 8\pi^2 \sigma \,, \quad g_- = \frac{1}{3\sigma} \,,
\end{equation}
as stated in the main text. We can regard these values of $g_{\pm}$ as gauge couplings at some fixed intermediate energy scale. Under RG flow, it can be shown (following the treatment in Refs.~\cite{Zou2020_DMITbilayer,Zou2020_DMITU2}) that the couplings obey the following leading-order equations
\begin{equation}
    \frac{d g_+}{dl} = - (g_+ + g_-) g_+ \,, \quad \frac{d g_-}{dl} = - (g_+ + g_-) g_- \,, \quad \frac{dV_{{\rm BCS}, m}}{dl} = g_+ - g_- - V_{{\rm BCS}, m}^2 \,,
\end{equation}
where $l$ is a running length scale and $V_{{\rm BCS}, m}$ is the BCS interaction in angular momentum channel $m$. From these flow equations, one can infer that the dimensionless ratio $g_+/g_-$ is an RG invariant. Thus, if $g_+ > g_-$ at some energy scale, the inequality persists down to the zero-temperature limit. This is an important fact that we use to establish the stability of MGQCPs in the main text. 

\subsection{Unscreened Coulomb interaction: $\epsilon = 0$}

The special case of unscreened Coulomb interactions is more subtle. In this case, $\alpha_-^0$ remains screened by the Fermi surface fluctuations and can be integrated to generate a renormalization of the coefficient $C$. However, integrating out $\alpha_+^0$ generates a $|\bs{q}|$-linear kinetic term for $\alpha_+^T$ which competes with the CFT contribution. Putting together all relevant terms, we find the low energy Euclidean Lagrangian
\begin{equation}
    L_{\rm eff}[\alpha_i] \approx \frac{1}{2} \left[\frac{|\bs{q}|}{g_+} + \frac{\gamma |\omega|}{|\bs{q}|}\right] \, |\alpha_+^T|^2 + \frac{1}{2} \left[\frac{|\bs{q}|}{g_-} + \frac{\gamma |\omega|}{|\bs{q}|}\right] \, |\alpha_-^T|^2 \,,
\end{equation}
where the positive constants $g_{\pm}$ are given by
\begin{equation}
    g_+^{-1} = \frac{1}{48 \pi^2 \sigma} + \frac{1}{16\pi^2 \left[\frac{1}{48 \pi^2 \sigma} + \frac{1}{2 A_0}\right]} \,, \quad g_-^{-1} = 3 \sigma \,, \quad A_0 = 2 \pi V_0 \,. 
\end{equation}
Since the ratio $g_+/g_-$ is preserved by the RG flow, we conclude that when $g_+ > g_-$ at a fixed energy scale, the repulsion mediated by $\alpha_+$ dominates over the attraction mediated by $\alpha_-$ down to the lowest energy scales.

The criterion $g_+ > g_-$ translates to an upper bound on the bare Coulomb interaction strength $V_0$
\begin{equation}
    V_0 < 12 \pi \sigma \left(144 \pi^2 \sigma^2 - 1\right) \,, \quad \forall \sigma > \frac{1}{12 \pi} \,. 
\end{equation}
Therefore, whenever $\sigma > 1/(12\pi)$, there is always an open neighborhood in parameter space where the critical Fermi surfaces survive. However, whether this criterion is sensible in any specific model depends on microscopic details. Restoring units, we see that 
\begin{equation}
    V_0 = \frac{e^2}{4 \pi \epsilon_0 \epsilon_r \hbar v_D} = \frac{\alpha c}{\epsilon_r v_D} \,, 
\end{equation}
where $\alpha$ is the fine-structure constant, $\epsilon_r$ is the dielectric screening factor, $c$ is the speed of light, and $v_D$ is the velocity of each gapless Dirac cone. Therefore, we obtain a useful dimensionless lower bound
\begin{equation}
    \frac{\epsilon_r v_D}{c} > \frac{\alpha}{12 \pi \sigma (144 \pi^2 \sigma^2 - 1)} \,. 
\end{equation}
Plugging in the infinite-$N_f$ conductivity $\sigma \approx 1/16$ (which is a reasonable estimate at $N_f = 6$), the bound simplifies to
\begin{equation}
    \frac{\epsilon_r v_D}{c} \gtrsim 6.8 \times 10^{-4} \,. 
\end{equation}
For any specific material, the dimensionless ratio $\epsilon_r v_D/c$ can be estimated and compared with the bound above.

\section{Pairing instability of generalized bilayer composite Fermi liquids}\label{app:CFL_pairing}

In this section, we provide some technical details on the interlayer pairing instability of two generalized CFL bilayers that appear on two sides of the phase transition described in the main text. For maximal generality, we will work at a layer filling fraction $\nu_i = \frac{p}{2p+1}$ where $p$ is an arbitrary integer (the minimal example in the main text corresponds to $p = 1$). The general effective Lagrangian takes the form
\begin{equation}
    L_{\rm eff} = \sum_{i=1}^2 L[f_i, A_i - \alpha_i] + \sum_{i=1}^2 L[\Phi_i, \alpha_i] \,. 
\end{equation}
For general $p$, we fractionalize $\Phi_i$ as $\Phi_i = \chi_i \psi$ and introduce a new $U(1)$ gauge field $a$ such that
\begin{equation}
    \sum_{i=1}^2 L[\Phi_i, \alpha_i] = \sum_{i=1}^2 L[\chi_i, \alpha_i - a] + L[\psi, a] \,.
\end{equation}

\subsection{Pairing instability of gCFL-A}

At filling $\nu_{\Phi_i} = p/(2p+1)$, the minimal topological order is the bFCI-A phase described in Appendix~\ref{app:largeNf}
\begin{equation}
    L^{(A)}_{{\rm bFCI}, p}[\Phi_i, \alpha_i] = \frac{2p+1}{4\pi} a da - \frac{p}{2\pi} a d (\alpha_1 + \alpha_2) + \frac{p}{4\pi} \alpha_1 d \alpha_1 + \frac{p}{4\pi} \alpha_2 d \alpha_2 \,. 
\end{equation}
Coupling this bosonic FCI back to the $\{f_i\}$ Fermi surfaces, we find the generalized CFL Lagrangian on the A-side of the phase transition
\begin{equation}
    L^{(A)}_{{\rm gCFL}, p} = \sum_{i=1}^2 L[f_i, A_i - \alpha_i] + \frac{2p+1}{4\pi} a da - \frac{p}{2\pi} a d (\alpha_1 + \alpha_2) + \frac{p}{4\pi} \alpha_1 d \alpha_1 + \frac{p}{4\pi} \alpha_2 d \alpha_2 \,. 
\end{equation}
To analyze the pairing instability of this gCFL, we can neglect subtleties of Chern-Simons level quantization and integrate out the gauge field $a$. The resulting Lagrangian simplifies to
\begin{equation}
     L^{(A)}_{{\rm gCFL}, p} = \sum_{i=1}^2 L[f_i, A_i - \alpha_i] + \frac{p}{4\pi} \alpha_1 d \alpha_1 + \frac{p}{4\pi} \alpha_2 d \alpha_2 - \frac{p^2}{2p+1} \frac{1}{4\pi} (\alpha_1 + \alpha_2) d (\alpha_1 + \alpha_2) \,. 
\end{equation}
The equations of motion for $\alpha_1, \alpha_2$ implement the generalized flux attachment constraints
\begin{equation}
    \rho_{f_i} = \frac{p}{2\pi} \nabla \times \bs{\alpha}_i - \frac{p^2}{2p+1} \frac{1}{2\pi} \nabla \times (\bs{\alpha}_1 + \bs{\alpha}_2) \,.
\end{equation}
Exploiting the $\mathbb{Z}_2$ exchange symmetry, we can define 
\begin{equation}
    \rho_{\pm} = \frac{1}{\sqrt{2}} (\rho_{f_1} \pm \rho_{f_2}) \,, \quad \alpha_{\pm} = \frac{1}{\sqrt{2}} (\alpha_1 \pm \alpha_2) \,, 
\end{equation}
so that the constraints simplify to 
\begin{equation}
    \rho_+ = \frac{1}{2\pi} \frac{p}{2p+1} \nabla \times \bs{\alpha}_+ \,, \quad \rho_- = \frac{p}{2\pi} \nabla \times \bs{\alpha_-} \,. 
\end{equation}
Through this flux attachment constraint, we can rewrite the Coulomb interaction matrix as 
\begin{equation}
    \begin{aligned}
    L_{\rm int} &\equiv \frac{1}{2} \begin{pmatrix}
        \rho_{f_1} & \rho_{f_2}
    \end{pmatrix} \begin{pmatrix}
        V & U \\ U & V 
    \end{pmatrix} \begin{pmatrix}
        \rho_{f_1} \\ \rho_{f_2}
    \end{pmatrix} = \frac{1}{2} \left[V(\bs{q}) + U(\bs{q})\right] |\rho_+(\bs{q})|^2 + \frac{1}{2} \left[V(\bs{q}) - U(\bs{q})\right] |\rho_-(\bs{q})|^2 \\
    &= \frac{1}{8\pi^2} \left(\frac{p}{2p+1}\right)^2 \left[V(\bs{q}) + U(\bs{q})\right]\, |\nabla \times \bs{\alpha}_+(\bs{q})|^2 + \frac{p^2}{8\pi^2} \left[V(\bs{q}) - U(\bs{q})\right]\, |\nabla \times \bs{\alpha}_-(\bs{q})|^2 \,.
    \end{aligned}
\end{equation}
The properties of this effective gauge field Lagrangian depend on the form of interactions. For the general class of interactions analyzed in Appendix~\ref{app:tunable_interaction}
\begin{equation}
    V(\bs{r}) = \frac{V_0}{|\bs{r}|^{1+\epsilon}} \,, \quad U(\bs{r}) = \frac{V_0}{\sqrt{|\bs{r}|^2+d^2}^{(1+\epsilon)}} \,,
\end{equation}
we see that $V(\bs{q}) + U(\bs{q})$ scales as $|\bs{q}|^{\epsilon-1}$, while $V(\bs{q}) - U(\bs{q})$ scales as $|\bs{q}|^0$ in the small $\bs{q}$ limit for all $\epsilon \geq 0$. This means that for all values of $p$, as long as $0 \leq \epsilon < 1$, generalized flux attachment keeps the fluctuations of $\alpha_+$ stiff, while softening the fluctuations of $\alpha_-$. The resulting RG flow takes the form (see Ref.~\cite{Zou2020_DMITbilayer,Zou2020_DMITU2})
\begin{equation}
    \frac{d g_+}{dl} = \frac{\epsilon}{2} \, g_+ - (g_++g_-) g_+ \,, \quad \frac{d g_-}{dl} = \frac{1}{2} \, g_- - (g_++g_-) g_-\,, \quad \frac{dV_{{\rm BCS}, m}}{dl} = g_+ - g_- - V_{{\rm BCS}, m}^2 \,. 
\end{equation}
The only stable fixed point of the $g_{\pm}$ flow equations is $(g_+ = 0, g_- = 1/2)$. At this fixed point, the BCS couplings $V_{{\rm BCS}, m}$ flow towards negative infinity. The physical interpretation is that the attraction mediated by $\alpha_-$ dominates over the repulsion mediated by $\alpha_+$, thereby inducing pairing of $f_1$ and $f_2$. 

For the marginal case $\epsilon = 1$, the effective kinetic terms for $\alpha_+^T$ is proportional to $|\bs{q}|^2 \log (1/|\bs{q}|)$, which is logarithmically enhanced relative to the $|\bs{q}|^2$ kinetic term of $\alpha_-^T$. As a result, the pairing instability still occurs, albeit at a lower temperature scale relative to the $\epsilon < 1$ case. 

When $\epsilon > 1$ (e.g. the standard gate-screened Coulomb interactions), the effective kinetic terms for both $\alpha_{\pm}^T$ are proportional to $|\bs{q}|^2$, with non-universal prefactors. Under RG flow, we have 
\begin{equation}
    \frac{d g_+}{dl} = \frac{1}{2} \, g_+ - (g_++g_-) g_+ \,, \quad \frac{d g_-}{dl} = \frac{1}{2} \, g_- - (g_++g_-) g_-\,, \quad \frac{dV_{{\rm BCS}, m}}{dl} = g_+ - g_- - V_{{\rm BCS}, m}^2 \,. 
\end{equation}
The flow of $V_{{\rm BCS}, m}$ under these equations depends on the choice of initial conditions. Whenever $g_+ < g_-$ holds at some intermediate energy scale, the BCS couplings flow towards $-\infty$ in the IR limit. Therefore, pairing still occurs in an open region of the microscopic parameter space. 

\subsection{Pairing instability of gCFL-B}

Next, we consider the B-side of the phase diagram, which can be accessed through a band-inversion transition with $\Delta C_{\chi_i} = - (2p+1)$. The resulting bosonic FCI is described by the TQFT
\begin{equation}
     L^{(B)}_{{\rm bFCI}, p}[\Phi_i, \alpha_i] = - \frac{2p+1}{4\pi} a da + \frac{p+1}{2\pi} a d (\alpha_1 + \alpha_2) - \frac{p+1}{4\pi} \alpha_1 d \alpha_1 - \frac{p+1}{4\pi} \alpha_2 d \alpha_2 \,.
\end{equation}
Including $\{f_i\}$ and integrating out $a$ now generates
\begin{equation}
    L^{(B)}_{{\rm gCFL}, p} = \sum_{i=1}^2 L[f_i, A_i - \alpha_i] - \frac{p+1}{4\pi} \alpha_1 d \alpha_1 - \frac{p+1}{4\pi} \alpha_2 d \alpha_2 + \frac{(p+1)^2}{2p+1} \frac{1}{4\pi} (\alpha_1 + \alpha_2) d (\alpha_1 + \alpha_2) \,. 
\end{equation}
The flux attachment constraints in the symmetric-antisymmetric basis now take the form
\begin{equation}
    \rho_+ = \frac{p+1}{2p+1} \frac{1}{2\pi} \nabla \times \bs{\alpha}_+ \,, \quad \rho_- = - \frac{p+1}{2\pi} \nabla \times \bs{\alpha}_- \,. 
\end{equation}
In the same basis, the density-density interaction reduces to
\begin{equation}
    H_{\rm int} = \frac{1}{8\pi^2} \left(\frac{p+1}{2p+1}\right)^2 \left[V(\bs{q}) + U(\bs{q})\right]\, |\nabla \times \bs{\alpha}_+(\bs{q})|^2 + \frac{(p+1)^2}{8\pi^2} \left[V(\bs{q}) - U(\bs{q})\right]\, |\nabla \times \bs{\alpha}_-(\bs{q})|^2 \,.
\end{equation}
For all $p \neq -1$, flux attachment again softens $\alpha_-$ while keeping $\alpha_+$ stiff, leading to a robust interlayer pairing instability for $0 \leq \epsilon < 1$. For $\epsilon \geq 1$, pairing does not always occur but remains the leading instability in an open region of the microscopic coupling constants. 

\section{An avoided MGQCP}\label{app:psi_sector_transition}

In this section, we consider another natural choice of bilayer bosonic FCI (which we refer to as bFCI-C) for $\{\Phi_i\}$ that can be accessed from bFCI-A through a direct continuous transition. The upshot is that the pairing instability of the $\{f_i\}$ Fermi surfaces persist at the critical point and becomes invisible at the lowest temperature. 

Recall that bFCI-A is constructed by writing $\Phi_i = \chi_i \psi$ and putting $\chi_i, \psi$ in IQH states with Hall conductance $\sigma_{xy} = 1$. To obtain a continuous bandwidth-tuned transition to a neighboring phase, we can make the Chern number of $\chi_i$ jump by $\pm 3$, or the Chern number of $\psi$ jump by $\pm 3$. In the main text, we chose the former possibility. Here, we instead consider the latter possibility and choose $C_{\psi}$ to jump from 1 to 4. The resulting bFCI-$\psi$ is described by a new TQFT
\begin{equation}
    L_{\rm bFCI}^{(C)} = \frac{4}{4\pi} a da + \frac{1}{4\pi} (\alpha_1 - a) d (\alpha_1 - a) + \frac{1}{4\pi} (\alpha_2 - a) d (\alpha_2 - a) \,. 
\end{equation}
Following the same procedure as in Appendix~\ref{app:CFL_pairing}, we can show that the Coulomb interactions reduce to
\begin{equation}
    L_{\rm int} = \frac{25(V + U)}{288\pi^2} |\nabla \times \bs{\alpha}_+|^2 + \frac{V-U}{8\pi^2} |\nabla \times \bs{\alpha}_-|^2 \,. 
\end{equation}
Therefore, bFCI-$\psi$ continues to induce a pairing instability between $f_1$ and $f_2$. 

At the critical point, the bosonic sector is described by the CFT 
\begin{equation}
    L_{\rm crit} = \sum_{n=1}^3 L[\Psi_n, a] + \frac{5}{2} \frac{1}{4\pi} a da + \frac{1}{4\pi} (\alpha_1 - a) d (\alpha_1 - a) + \frac{1}{4\pi} (\alpha_2 - a) d (\alpha_2 - a) \,,
\end{equation}
where $\Psi_n$ is a gapless Dirac fermion regularized to have vanishing Hall conductance. From the form of the CFT Lagrangian, it is clear that the gapless fluctuations of $\Psi_n$ only couple to the symmetric gauge field $\alpha_+$ and not to the antisymmetric gauge field $\alpha_-$. Therefore, the effective gauge field kinetic term generated by the CFT takes the schematic form
\begin{equation}
    L_{\rm crit}[\alpha_i] = \frac{1}{2} \alpha_+^{\mu} (A \, q\, P_{\mu\nu}) \alpha_+^{\nu} + \textrm{Chern-Simons terms} \,. 
\end{equation}
Without computing the coefficient $A$, one can immediately infer that $\alpha_+$ stiffens at the critical point, while $\alpha_-$ remains soft with a $|\bs{q}|^2$ dispersion generated by diamagnetic Fermi surface fluctuations. Hence, singular $\alpha_-$ fluctuations mediate an attraction between $f_1$ and $f_2$ which dominates over the repulsion mediated by $\alpha_+$, leading to a pairing instability.

Assuming that the Fermi surface is always paired in a fixed angular momentum channel $l$, we can simply integrate out the fermions $f_i$ to generate a low energy effective Lagrangian with no Fermi surfaces 
\begin{equation}
    L_{\rm eff} = \frac{l}{4\pi} (A_1 - \alpha_1) d (A_1 - \alpha_1) + \sum_{n=1}^3 L[\Psi_n, a] + \frac{5}{2} \frac{1}{4\pi} a da + \frac{1}{4\pi} (\alpha_1 - a) d (\alpha_1 - a) + \frac{1}{4\pi} (A_1 + A_2 - \alpha_1 - a) d (A_1 + A_2 - \alpha_1 - a) \,. 
\end{equation}
This Lagrangian describes a conventional CFT that separates two gapped fermionic FCIs. 

\section{Two classes of generalizations}

In this section, we consider two simple generalizations of the minimal MGQCP analyzed in the main text. Both cases have interesting features that are not present in the minimal example, though the basic phase diagram/crossover structure remains unchanged. 

\subsection{Opposite Chern number jumps}\label{app:opposite_Chernjump}

We first consider an obvious generalization in which the bosonic sector undergoes a different topological transition. Recall that $\Phi_i$ is at layer filling $\nu_i = 1/3$. Within the parton decomposition $\Phi_i = \chi_i \psi$, bFCI-A is constructed by choosing the flux assignment such that $\chi_i, \psi$ form gapped Chern insulators with $C_{\chi_i} = 1 = C_{\psi} = 1$. The resulting TQFT is
\begin{equation}
    L_{\rm bFCI}^{(A)} = \frac{1}{4\pi} (\alpha_1 - a) d (\alpha_1 - a) + \frac{1}{4\pi} (\alpha_2 - a) d (\alpha_2 - a) + \frac{1}{4\pi} a da + 6 \mathrm{CS}_g \,.  
\end{equation}
In the main text, we considered a transition across which $C_{\chi_i}$ jumps to $-2$. An obvious alternative is for $C_{\chi_i}$ to jump to $4$. If we consider this possibility, then the bosonic sector transitions to a distinct bFCI-C described by a TQFT
\begin{equation}
    L_{\rm b-FCI}^{(C)} = \frac{4}{4\pi} (\alpha_1 - a) d (\alpha_1 - a) + \frac{4}{4\pi} (\alpha_2 - a) d (\alpha_2 - a) + \frac{1}{4\pi} a da + 18 \mathrm{CS}_g \,. 
\end{equation}
The transition between these two bFCIs is again $N_f = 6$ QED$_3$, but with a Chern-Simons term at level $k = 6$
\begin{equation}
    L_{\rm crit,\Phi}^{(A \leftrightarrow C)} = \sum_{i=1}^2 \sum_{n = 1}^3 L[\Psi_{i, n}, \alpha_i - a] + \frac{1}{4\pi} a da + \frac{5}{2} \frac{1}{4\pi} (\alpha_1 - a) d (\alpha_1 - a) + \frac{5}{2} \frac{1}{4\pi} (\alpha_2 - a) d (\alpha_2 - a) \,.  
\end{equation}
This critical theory belongs to a general class of theories 
\begin{equation}
    L[k_a, k_{\alpha}] = \sum_{i=1}^2 \sum_{n = 1}^3 L[\Psi_{i, n}, \alpha_i - a] + \frac{k_a}{2} a E a + \frac{k_{\alpha}}{2} (\alpha_1 - a) E (\alpha_1 - a) + \frac{k_{\alpha}}{2} (\alpha_2 - a) E (\alpha_2 - a) \,,
\end{equation}
where $E$ is the Chern-Simons operator $E^{\mu\nu} = \frac{1}{2\pi} \epsilon^{\mu\lambda\nu} \partial_{\lambda}$. Assuming a zero-temperature Dirac fermion conductivity $\sigma q P_{\mu\nu}$ where $P$ is the spacetime transverse projector, we can integrate out the Dirac fermions and $a$ to find an approximate effective Lagrangian for $\alpha_i$
\begin{equation}
    L_{\rm crit, \Phi}[\alpha_i, k_a, k_{\alpha}] = \frac{1}{2} \alpha_+^{\mu} K_{+,\mu\nu} \alpha_+^{\nu} + \frac{1}{2} \alpha_-^{\mu} K_{-,\mu\nu} \alpha_-^{\nu} \,, 
\end{equation}
where the kernels $K_+$ and $K_-$ are defined by
\begin{equation}
    \begin{aligned}
    K_{+, \mu\nu} &= 3 \sigma q \frac{\frac{k_a^2}{4\pi^2}}{\frac{(k_a + 2 k_{\alpha})^2}{4\pi^2} + 36 \sigma^2} P_{\mu\nu} + \frac{k_a \left[18 \sigma^2+ \frac{k_{\alpha} (k_a + 2 k_{\alpha})}{4\pi^2}\right]}{\frac{(k_a + 2 k_{\alpha})^2}{4\pi^2} + 36 \sigma^2} E_{\mu\nu} \,,\\ 
    K_{-, \mu\nu} &= 3 \sigma q P_{\mu\nu} + k_{\alpha} E_{\mu\nu}  \,.
    \end{aligned}
\end{equation}
Next, we incorporate Fermi surface effects. For simplicity, we will consider screened Coulomb interactions with $\epsilon > 0$ (see Appendix~\ref{app:tunable_interaction} for a treatment of the unscreened case). In this case, Fermi surface fluctuations suppress the $\alpha_{\pm}^0$ fluctuations such that all terms involving $\alpha_{\pm}^0$ can be dropped in the low energy limit. Moreover, Fermi surface fluctuations give rise to a diamagnetic $|\bs{q}|^2$ term and a Landau damping $\omega/|\bs{q}|$ term in the kinetic term of $\alpha_{\pm}^T$. After incorporating all of these effects, we find the Euclidean Lagrangian
\begin{equation}
    L_{\rm crit, \Phi}[\alpha_i,k_a,k_{\alpha}] \approx \frac{1}{2} \left[C \, |\bs{q}|^2  + \gamma |\omega|/|\bs{q}| + 3 \sigma |\bs{q}| \frac{k_a^2}{(k_a + 2k_{\alpha})^2 + 144\pi^2 \sigma^2}\right] \, |\alpha_+^T|^2 + \frac{1}{2} \left[C \, |\bs{q}|^2  + \gamma |\omega|/|\bs{q}| + 3 \sigma |\bs{q}|\right] \,|\alpha_-^T|^2 \,. 
\end{equation}
As a sanity check, note that choosing $k_a = 1, k_{\alpha} = -1/2$ recovers \eqref{eq:renormalized_alpha} in the main text. Choosing instead $k_a = 1, k_{\alpha} = \frac{5}{2}$, we find 
\begin{equation}
    L_{\rm crit, \Phi}[\alpha_i,k_a,k_{\alpha}] \approx \frac{1}{2} \left[C \, |\bs{q}|^2  + \gamma |\omega|/|\bs{q}| + 3 \sigma |\bs{q}| \frac{1}{36 + 144\pi^2 \sigma^2}\right] \, |\alpha_+^T|^2 + \frac{1}{2} \left[C \, |\bs{q}|^2  + \gamma |\omega|/|\bs{q}| + 3 \sigma |\bs{q}|\right] \, |\alpha_-^T|^2 \,. 
\end{equation}
More generally, it is straightforward to check that the coefficient of $|\bs{q}|$ in the $\alpha_+^T$ kinetic term is always smaller than the coefficient of $|\bs{q}|$ in the $\alpha_-^T$ kinetic term whenever $|k_a + 2 k_{\alpha}| \geq |k_a|$. Under these conditions, the $\alpha_+$ fluctuations are always softer \textit{independent of the value of $\sigma$} and a pairing instability between $f_1, f_2$ is averted. 

To fully establish the stability of the MGQCP between FCI-A and FCI-C, one final condition we need to check is that the correlation length exponent $\nu$ of the $N_f = 6, k = 6$ QED$_3$-CS theory satisfies the bound $\nu > 2/3$. Using a large-$N_f$ expansion with $\kappa = k/N_f$ fixed, we have the following result for the correlation length exponent
\begin{equation}
    \nu^{-1}(N_f, \kappa) = 1 + \frac{1}{N_f} \frac{512}{3 \pi^2} \frac{\phi (1-2\phi)}{(1 + \phi)^3} + \mathcal{O}\left(\frac{1}{N_f^2}\right) \,, \quad \phi = \left(\frac{\pi}{8\kappa}\right)^2 \,. 
\end{equation}
Plugging in $N_f = 6$ and $\kappa = 1$, we find
\begin{equation}
    \nu(6, 1) \approx 0.833 > 2/3 \,. 
\end{equation}
Therefore, at leading two orders in the large-$N_f$ expansion, the CFT sector decouples from the Fermi surface sector and the phenomenology of the MGQCP between FCI-A and FCI-C remains qualitatively identical to the MGQCP between FCI-A and FCI-B.

\subsection{General filling fraction}\label{app:largeNf}

As another natural generalization, we consider the electron bilayer at filling $\nu_i = p/(2p+1)$ with $p \in \mathbb{Z}$. Following the same procedure as in the main text, we fractionalize the electron as $c_i = f_i \Phi_i$ with $\Phi_i = \chi_i \psi$ and assign all the magnetic flux to $\Phi_i$. The minimal topological order for $\{\Phi_i\}$ at this filling can be constructed through a mean-field ansatz $\nabla \times \bs{a} = 2p B/(2p+1)$. Since $\psi$ is at Landau level filling 1 and $\chi_i$ is at Landau level filling $p$, the fermionic partons $\psi$ and $\chi_i$ can go into Chern insulator states with $C_{\psi} = 1$ and $C_{\chi_i} = p$ respectively. The resulting TQFT takes the form
\begin{equation}
    L_{{\rm bFCI}, p}^{(A)}[\Phi_i, \alpha_i] = \frac{p}{4\pi} (\alpha_1 - a) d (\alpha_1 - a) + \frac{p}{4\pi} (\alpha_2 - a) d (\alpha_2 - a) + \frac{1}{4\pi} a da \,. 
\end{equation}
Across a bandwidth-tuned transition, the Chern number of each fermionic parton can jump by multiples of $|2p+1|$, as each fermionic parton sees fractional flux $1/(2p+1)$. Following the strategy in the main text, we will consider a transition in which $C_{\chi_i}$ jumps from $p$ to $-(p+1)$. The new TQFT takes the form
\begin{equation}
    L_{{\rm bFCI}, p}^{(B)}[\Phi_i, \alpha_i] = - \frac{p+1}{4\pi} (\alpha_1 - a) d (\alpha_1 - a) - \frac{p+1}{4\pi} (\alpha_2 - a) d (\alpha_2 - a) + \frac{1}{4\pi} a da \,. 
\end{equation}
The critical theory between these two bFCIs is 
\begin{equation}
    L_{{\rm crit}, p}^{(A \leftrightarrow B)}[\Phi_i, \alpha_i] = \sum_{i=1}^2 \sum_{n = 1}^{|2p+1|} L[\Psi_{i,n}, \alpha_i - a] + \frac{1}{4\pi} a da - \frac{1}{8\pi} (\alpha_1 - a) d (\alpha_1 - a) - \frac{1}{8\pi} (\alpha_2 - a) d (\alpha_2 - a) \,. 
\end{equation}
Compared with the bosonic CFT \eqref{eq:bFCI_CFT} in the main text, we see that this new critical theory is again QED$_3$ but with $N_f = 2|2p+1|$. 

Alternatively, following the procedure in Appendix~\ref{app:opposite_Chernjump}, we can postulate the opposite Chern number jump $C_{\chi_i} \rightarrow 3p + 1$. Under this choice, we obtain a distinct bFCI-C described by the Lagrangian 
\begin{equation}
    L_{{\rm bFCI}, p}^{(C)} = \frac{3p+1}{4\pi} (\alpha_1 - a) d (\alpha_1 - a) + \frac{3p+1}{4\pi} (\alpha_2 - a) d (\alpha_2 - a) + \frac{1}{4\pi} a da \,. 
\end{equation}
The critical theory between bFCI-A and bFCI-C is a QED$_3$-CS theory with $N_f = 2|2p+1|$ and $k = 2(2p+1)$
\begin{equation}
    L_{{\rm crit},p}^{(A \leftrightarrow C)}[\Phi_i, \alpha_i] = \sum_{i=1}^2 \sum_{n=1}^{|2p+1|} L[\Psi_{i,n}, \alpha_i - a] + \frac{4p+1}{2} \frac{1}{4\pi} (\alpha_1 - a) d (\alpha_1 - a) + \frac{4p+1}{2} \frac{1}{4\pi} (\alpha_2 - a) d (\alpha_2 - a) + \frac{1}{4\pi} a da \,. 
\end{equation}
As $p \rightarrow \infty$, both $L_{{\rm crit},p}^{(A \leftrightarrow B)}$ and $L_{{\rm crit},p}^{(A \leftrightarrow C)}$ analytically controlled CFTs with $\nu > 2/3$. We therefore obtain a two-parameter family of analytically controlled MGQCPs.

\section{Evolution of anyon energy gaps in the approach to criticality}\label{app:gap_scaling}

In the main text, we claimed that the pairing of generalized CFLs in the vicinity of the MGQCP creates an asymptotic separation between the energy gaps of different anyons. The goal of this section is to justify this claim for gCFL-A, following ideas in Ref.~\cite{Metlitski2014_NFLpairing}. The argument for gCFL-B is the same up to unimportant constants.

Recall that below the energy scale $\Delta_{\Phi}^2$, the gCFL-A is well-described by the effective Lagrangian 
\begin{equation}
    L_{\rm gCFL}^{(A)} = \sum_{i=1}^2 L[f_i, A_i - \alpha_i] + 3 \mathrm{CS}[a,g] - \frac{1}{2\pi} a d (\alpha_1 + \alpha_2) + \frac{1}{4\pi} \alpha_1 d \alpha_1 + \frac{1}{4\pi} \alpha_2 d \alpha_2 \,. 
\end{equation}
We can regard $\Delta_{\Phi}^2$ as a UV cutoff as far as the $\{f_i\}$ fermions are concerned. From the analysis in Ref.~\cite{Metlitski2014_NFLpairing}, we know that pairing occurs at a parametrically smaller energy scale $\Delta_f \sim c \Delta_{\Phi}^2$, where $c$ is a non-universal constant. Therefore, the anyon associated with the BdG quasiparticle has an energy gap set by $\Delta_f$. 

Near the critical point, we can define a fermion correlation length $\xi_f$ which is related to the fermion energy gap through $\xi_f \sim \Delta_f^{-1/z_f}$, with $z_f$ the fermion dynamical exponent in the Fermi surface sector. Following Ref.~\cite{Metlitski2014_NFLpairing}, we model a vortex by replacing a region of size $\xi_f^2$ with the unpaired normal state which competes with the paired ground state. The vortex gap thus scales as $\Delta_v \sim \xi_f^2 (\epsilon_n - \epsilon_p)$ where $\epsilon_n - \epsilon_p$ is the difference in energy density between the normal state and the paired state. Using $\xi_f \sim \Delta_f^{-1/z_f}$ and $\epsilon_n - \epsilon_p \sim \Delta_f^{1+1/z_f}$, we conclude that $\Delta_v \sim \Delta_f^{1 - 1/z_f}$. 

The parent normal state of FCI-A is gCFL-A, which has a gauge field dynamical exponent $z_- = 3$ for $\alpha_-$. As a result, the fermion dynamical exponent is $z_f = 3/2$ and we arrive at
\begin{equation}
    \Delta_v \sim \Delta_f^{-2/z_f} \times \Delta_f^{1 + 1/z_f} \sim \Delta_f^{1 - 1/z_f} \sim \Delta_f^{1/3} \sim \Delta_{\Phi}^{2/3} \gg \Delta_{\Phi} \,,
\end{equation}
which is precisely the hierarchy of energy scales in the main text.

\end{document}

%% file: preamble.tex
\usepackage{graphicx}
\usepackage{dcolumn}
\usepackage{bm}
\usepackage{multirow}

\usepackage[normalem]{ulem}

\usepackage{amsmath}
\usepackage{amsthm}
\usepackage{amstext}
\usepackage{amssymb}
\usepackage{mathrsfs}
\usepackage{amsfonts}
\usepackage{amsbsy} 

\usepackage{braket}

\usepackage[all,matrix,cmtip]{xy}
\usepackage{mathtools}

\usepackage{xcolor}
\definecolor{red}{rgb}{1,0,0}
\definecolor{blue}{rgb}{0,0,1}
\definecolor{dblue}{rgb}{0,0,0.4}
\definecolor{green}{rgb}{0,1,0}
\definecolor{black}{rgb}{0,0,0}
\definecolor{white}{rgb}{1,1,1}
\definecolor{pastelblue}{RGB}{20,93,160}

\definecolor{brn}{rgb}{.8,.4,.0}
\definecolor{redo}{rgb}{1,.5,.0}
\definecolor{ddgrn}{rgb}{0,0.4,0}
\definecolor{dgrn}{rgb}{0,0.55,0}
\definecolor{dbl}{rgb}{0,0,0.5}

\usepackage[colorlinks,citecolor=pastelblue,linkcolor=pastelblue,urlcolor=pastelblue]{hyperref}

\newcommand{\sgn}{{\rm sgn}}

\newcommand{\bpm}{\begin{pmatrix}}
	\newcommand{\epm}{\end{pmatrix}}
\newcommand{\bmm}{\begin{matrix}}
	\newcommand{\emm}{\end{matrix}}
\newcommand{\bvm}{\begin{vmatrix}}
	\newcommand{\evm}{\end{vmatrix}}

\usepackage{euscript}

\makeatletter
\newsavebox{\@brx}
\newcommand{\llangle}[1][]{\savebox{\@brx}{\(\m@th{#1\langle}\)}%
	\mathopen{\copy\@brx\kern-0.5\wd\@brx\usebox{\@brx}}}
\newcommand{\rrangle}[1][]{\savebox{\@brx}{\(\m@th{#1\rangle}\)}%
	\mathclose{\copy\@brx\kern-0.5\wd\@brx\usebox{\@brx}}}
\makeatother